\documentclass[12pt]{article}
\usepackage{latexsym, amsmath, epsfig, amsfonts, amsbsy}
\DeclareFontFamily{OT1}{rsfs10}{}
\DeclareFontShape{OT1}{rsfs10}{m}{n}{ <-> rsfs10 }{}
\DeclareMathAlphabet{\mathscript}{OT1}{rsfs10}{m}{n}

\else\oddsidemargin25mm\evensidemargin25mm\marginparwidth25mm\fi
\def\Z{\mathbb{Z}}
\def\C{\mathbb{C}}

\def\R{\mathbb{R}}

\def\N{\mathbb{N}}

\def\bpl{\Big(}
\def\bpr{\Big)}

\newcommand{\ft}[2]{{\textstyle\frac{#1}{#2}}}
\def\brr{\begin{equation}}
\def\err{\end{equation}}
\def\brr{\begin{eqnarray}}
\def\err{\end{eqnarray}}
\def\ba{\left(\begin{array}}
\def\ea{\end{array}\right)}
\def\lf{\left.\begin{array}{c}}
\def\rf{\end{array}\right.}

\newcommand{\dr}{\raise.3ex\hbox{$\stackrel{\leftarrow}{\partial }$}{}}
\newcommand{\dl}{\raise.3ex\hbox{$\stackrel{\rightarrow}{\partial}$}{}}
\newcommand{\topi}{\raise.3ex\hbox{$\stackrel{\pi}{\longrightarrow}$}{}}
\newcommand{\ns}{\normalsize}

\renewcommand{\a}{\alpha}
\renewcommand{\b}{\beta}
\newcommand{\g}{\gamma}
\newcommand{\s}{\sigma}
\newcommand{\G}{\Gamma}
\newcommand{\q}{\theta}
\begin{document}


\begin{titlepage}

\vspace{-4cm}

\title{
   \hfill{\ns CU-TP-1066\,,\,HWS-200201\\}
   \hfill{\ns hep-th/0208030\\[2cm]}
   {\LARGE A ``Periodic Table" for Supersymmetric {\it M}-Theory Compactifications \\[1cm]  }}

\author{{\bf
   Charles F. Doran$^{1}$ and
   Michael Faux$^{1,2}$}\\[5mm]
   {\it $^1$Departments of Mathematics and Physics} \\
   {\it Columbia University} \\
   {\it 2990 Broadway, New York, NY 10027} \\[3mm]
   {\it $^2$ Department of Physics} \\
   {\it Hobart and William Smith Colleges} \\
   {\it Geneva, NY 14456}}
\date{}

\maketitle

\vspace{.3in}

\begin{abstract}
\noindent
We develop a systematic method for classifying
supersymmetric orbifold compactifications of {\it M}-theory. By
restricting our attention to abelian orbifolds with low order, in
the special cases where elements do not include coordinate shifts,
we construct a ``periodic table" of such compactifications,
organized according to the orbifolding group (order $\leq 12$) and
dimension (up to $7$). An intriguing connection between
supersymmetric orbifolds and $G_2$-structures is explored.

\vspace{.5in} \noindent
\end{abstract}

\thispagestyle{empty}

\end{titlepage}


\section{Introduction}
 Manifolds with $SU(N)$ holonomy have been a source of significant
 interest for mathematicians and physicists alike.  Indeed, the
 importance of $K3$ manifolds and Calabi-Yau threefolds
 in the arena of consistent superstring background geometries could
 hardly be overstated.  Aside from their undisputed beauty,
 compactification on these spaces allows for
 important control of supersymmetry which, in turn, permits ready access
 to potentially testable phenomenological consequences of string
 theory itself.  It is for such reasons that understanding the geometry of these objects,
 including the classification of associated gauge bundles
 \cite{fmw1, fmw2, fmw3}, has been such a relevant and fruitful
 endeavor.
 It is nowadays accepted, however, that there exists a more
 fundamental eleven dimensional underpinning, code-named
 {\it M}-theory, which appropriately describes
 non-perturbative aspects of fundamental physics.
 In contrast to the situation in perturbative string theory,
 within the context of {\it M}-theory the most important geometric
 compactification spaces have special holonomy
 \cite{Ach, AtWit, GYZ}.

 The connection between {\it M}-theory and
 four-dimensional $N=1$ supersymmetric models of particle
 physics is provided by eleven-dimensional supergravity
 on compact seven-manifolds with $G_2$ holonomy.  Considerably
 less is known about these objects as compared to the case
 of Calabi-Yau manifolds.  In light of the above discussion,
 however, it is important to develop a useful classification
 of the relevant supersymmetric {\it M}-theory models.
 The rudiments of a mathematical
 classification scheme for $G_2$ holonomy seven manifolds, each a
 resolution of an orbifold of a seven-torus,
 has been provided by Joyce \cite{Joyce}.
 The purpose of this paper is to describe a complementary scheme,
 based on physics, of a class of seven-dimensional
 orbifold constructions which meet the criterion of $N=1$
 supersymmetry preservation.

 In previous papers \cite{mlo, phase, hetk3, DFO}
 we described various technical
 aspects of the extraction of effective physics from
 {\it M}-theory.  Generally, our techniques apply to global
 orbifold compactifications,
 and rely on significant constraints which follow from the requirement of chiral anomaly
 cancellation point-wise in eleven-dimensions, most notably on
 distinguished even-dimensional submanifolds.  Recently \cite{chiralz},
 we have described how to obtain a pair of particular four-dimensional
 $N=1$ super Yang-Mills theories with chiral matter content
 from an {\it M}-theoretic intersecting brane-world scenario.
 In that paper we included a scan of multiplicities of supersymmetric {\it M}-theory orbifold
 models of a particular class.  In this paper we derive
 this scan, explaining in more detail the physical and
 mathematical criteria involved in finding such models.

 Presupposing an ultimate connection between {\it M}-theory and standard model
 four-dimensional physics, a seven dimensional compactification
 space must be Ricci-flat and admit singularities
 \cite{CandRain}.  For these reasons, the class of toroidal orbifolds
 $T^7/\Gamma$, for a finite group $\Gamma$, holds special interest.
 The necessary geometrical singularities are of finite quotient type,
 and hence readily permit mathematical analysis. Each is modelled on
 $(M^1)^{7-n} \times {\mathbb{R}}^n/G$ for some subgroup $G \subset \Gamma$,
 where $M^1 = S^1$ or the unit interval $I^1 = S^1/{\mathbb{Z}}_2$.
 Moreover, as we shall see in this paper, under the right
 conditions one can explicitly describe a well-defined lift of the
 action of $\Gamma$ to the eleven-dimensional spinorial
 supercharge.  This allows us to determine how much supersymmetry
 is preserved on the various fixed-point loci (``fixed-planes") of
 spacetime $T^7/\Gamma \times {\mathbb{R}}^{3,1}$.

 The mathematical problem of identifying candidate compactification
 spaces with supersymmetric fixed-planes is quite elegant, and
 divides neatly into four parts.
 First of all, we must decide on a class of tori to orbifold.  A torus
 $T^7$ is determined by a choice of a rank seven lattice $\Lambda
 \subset {\mathbb{R}}^7$.  Throughout this paper we will assume that
 the lattice has the form $\Lambda := A_1 \oplus A_2^3$, i.e., the
 direct sum of three copies of the usual hexagonal lattice in the
 complex plane with one copy of ${\mathbb{Z}}$.  More generally, our
 analysis applies to compactification on tori $T^n$ modelled on
 lattices $A_1^a \oplus A_2^b$ with
 $a+2\,b=n$, $1 \leq n \leq 7$.
 These are by no means the only lattices from which we could
 construct our tori.  In fact, a particularly interesting case,
 especially as regards the discussion in Section \ref{softsect} of this paper,
 is that of the irreducible lattice $A_7 \subset {\mathbb{R}}^7$,
 whose automorphism group is one of the maximal finite subgroups of
 the group $G_2$.  We choose to restrict attention to our
 particular class of decomposable lattices simply because it is
 both sufficiently general to subsume the orbifolds studied
 previously, and easy in this setting to describe the action on
 the eleven-dimensional supercharge in the Clifford algebra.

 The second step in identifying the desired supersymmetric orbifolds is to
 choose a particular class of groups $\Gamma$ acting on $T^7$.
 An action on the torus is an action on
 ${\mathbb{R}}^7$ that preserves the lattice $\Lambda$. We will
 consider only group actions which respect the decomposition of
 $\Lambda$ into direct summands. Thus, an element $g \in \Gamma$
 acts as $\exp(\,2\,\pi\,\imath\,\vec{f}\,)$,
 where $\vec{f} =(\,f_1\,,\,f_2\,,\,f_3\,,\,f_4)$, with
 $f_{1,2,3} \in \Z/6$ and $f_4\in\Z/2$.  In this way
 we define a class of representations in which each element acts
 by rotations in two-dimensional subplanes plus the possibility
 of a parity reversal on one real coordinate.  We call such actions
 ``pseudo-planar representations", and groups which
 admit such representations ``pseudo-planar groups".   This
 eliminates, for the time being, orbifolds constructed
 from non-abelian orbifolding groups, an omission we hope to
 rectify in future work.

 Having restricted ourselves to pseudo-planar groups, next we need
 to enumerate all possible actions.   For organizational
 purposes we wish to index the candidate orbifolds of $T^n$ by their
 orbifolding groups $\Gamma$.
 For reasons of bounding complexity, in this paper we restrict
 attention to finite abelian groups $\Gamma$ of order $\leq 12$.
 The method we use, explained in detail in Section \ref{repslong}, consists of
 classifying the $\Lambda$-compatible representations of $\Gamma$
 on ${\mathbb{R}}^7$, using the decomposition into irreducible
 characters to determine equivalence classes of group actions.  Properly
 taking into account the automorphisms of the group $\Gamma$
 allows us to distinguish inequivalent group actions, and a matrix
 formalism makes quick work of computing the dimensions of the fixed-planes
 corresponding to each element of $\Gamma$, and
 provides for a concise accounting of various geometric
 data associated with each orbifold.

 Finally, given this data, we must have a criterion for determining
 the exact amount of supersymmetry preserved on each fixed-plane.
 This fourth and final ingredient is provided by a systematic analysis of
 lifts of $\Gamma$ to actions on eleven dimensional spinors.
 The necessary properties of Clifford algebra are reviewed in
 Section \ref{susynotat}, and the criterion,
 our Supersymmetric Restriction Theorem, is summarized in Table
 \ref{table1}.  We apply this criterion to the full class of groups $\Gamma$
 acting compatibly on our lattice $\Lambda$.  In this way, we
 identify and classify the relatively small number of orbifolds
 $T^n/\Gamma$ which maintain some supersymmetry at all points,
 constructions we refer to as {\it supersymmetric orbifolds}.

 In the language of \cite{chiralz}, the orbifolds considered here
 are all {\it hard} orbifolds, i.e., there are no fixed-point-free
 coordinate ``shifts" in
 the $\Gamma$ action, since we act directly through a representation on
 the space ${\mathbb{R}}^7$ over $T^7 = {\mathbb{R}}^7 / \Lambda$.
 By contrast, Joyce's examples of orbifolds of $T^7$ admitting a
 resolution as a $G_2$ manifold \cite{Joyce} are all {\it soft} orbifolds.  In
 Section \ref{softsect} we use the first class of $G_2$ resolvable examples studied
 by Joyce \cite{JoyJDG1, JoyJDG2} as a launching point for a
 discussion of the relationship between supersymmetric orbifolds of
 $T^7$ and the notion of a $G_2$-structure (a weaker, necessary condition
 for the orbifold to admit a $G_2$ holonomy resolution).

 We have collected the results of our search, accounting for all
 pseudo-planar orbifold groups with low order, into a so-called
 ``Periodic Table" of orbifolds, which we include in this
 introduction as Table \ref{scan2}.
 \begin{table}
 \begin{center}
 \begin{tabular}{|c||c|c|c|c|c|c|c|}
 \hline
 &
 & \hspace{.3in}
 & \hspace{.3in}
 & \hspace{.3in}
 & \hspace{.3in}
 & \hspace{.3in}
 & \hspace{.3in} \\[-.1in]
 $\Gamma$  & 1 & 2 & 3 & 4 & 5 & 6 & 7 \\[.1in]
 \hline
 \hline
 &&&&&&&\\[-.1in]
 $\Z_2$                & $(1)^*$ & & & (4) & (5) &  & \\[.1in]
 \hline
 &&&&&&&\\[-.1in]
 $\Z_3$                &       & & & (4) &  &  &  \\[.1in]
 \hline
 &&&&&&&\\[-.1in]
 $\Z_4$                &       & & & (04) & (14) & (24) & (34)  \\[.1in]
 \hline
 &&&&&&&\\[-.1in]
 $\Z_2\times \Z_2$     &       & & &   & $(014)^*$ & (222) & (223) \\[.1in]
 \hline
 &&&&&&&\\[-.1in]
 $\Z_2\times \Z_3$     &       & & & (004) & $(140)^*$ & (222) & (322) \\[.1in]
 &&&&& (104) & & \\[.1in]
 \hline
 &&&&&&&\\[-.1in]
 $\Z_2\times \Z_4$     &       & & &   & $(00104)^*$ & $(10104)^*$ & $(20104)^*$ \\[.1in]
 &&&&&& (00222) & (00322) \\[.1in]
 &&&&&& (01122) & (01222) \\[.1in]
 &&&&&&& (10222) \\[.1in]
 &&&&&&& (11122) \\[.1in]
 \hline
 &&&&&&&\\[-.1in]
 $(\Z_2)^3$            &       & & &   &  &  & $(2220001)^*$ \\[.1in]
 &&&&&&& (1111111) \\[.1in]
 \hline
 &&&&&&&\\[-.1in]
 $(\Z_3)^2$            &       & & &   &  &  &  \\[.1in]
 \hline
 &&&&&&&\\[-.1in]
 $(\Z_2)^2\times \Z_3$ &      & & &   & $(1000004)^*$ & (0020220) & $(0122022)^*$ \\[.1in]
 &&&&&&& (0122020) \\[.1in]
 &&&&&&& (0030220) \\[.1in]
 \hline
 &&&&&&&\\[-.1in]
 $\Z_3\times \Z_4$     &      & & &   &       &   & \\[.1in]
 \hline
 \end{tabular} \\[.2in]
 \caption{The ``Periodic Table" listing all supersymmetric, hard,
 pseudo-planar abelian orbifolds
 $T^n/\Gamma$ of {\it M}-theory, for cases $| \Gamma | \le 12$.
 The labeling system is explained in Section \ref{repslong}.}
 \label{scan2}
 \end{center}
 \end{table}
 In Table \ref{scan2} we exhibit each supersymmetric orbifold
 $T^n/\Gamma$, with rows corresponding to distinct pseudo-planar
 groups $\Gamma$, listed by increasing group order, and
 columns corresponding to the representation dimensions $1\le n\le 7$.
 In each block of this table are listed the complete set of hard
 supersymmetric orbifolds corresponding to associated $n$-dimensional
 representations of $\Gamma$ compatible with our lattices.  Each
 orbifold is indicated by a particular {\it label}, which
 codifies the group action of $\Gamma$ on $T^n$ in a manner explained in
 detail in Section \ref{repslong}.  For a subset of the supersymmetric
 orbifolds, the corresponding orbifold label has an asterix
 appended.  These models are those which split off a separate
 $S^1/\Z_2$ factor.  Such models are the only ones which have
 ten-dimensional fixed-planes.   Owing to this distinction, there is
 a more direct connection
 between this class of supersymmetric orbifolds and perturbative
 heterotic string models than is the case for the orbifolds listed
 without stars.

 There are two natural extensions of our work in this paper which
 we plan to investigate in the near future. First,
 we would like to remove the pseudo-planar and abelian restrictions on
 the $\Gamma$ action,
 allowing instead any $\Gamma \subset \mbox{Aut}(\Lambda)$.
 In particular this
 will allow many nonabelian group actions, which in turn will
 require a generalization of the supersymmetric restriction
 proof of Section \ref{susypf}.  An important step towards such a
 formulation is described in Section \ref{nonabelian}.
 It would also be quite valuable to reformulate both the
 analysis herein and the anomaly cancellation compatibility
 checks in \cite{DFO, chiralz} using the theory of principal bundles on
 orbifolds.  In this setting both the supersymmetric restriction
 criterion and anomaly cancellation mechanism should find
 expression in the language of characteristic classes of such
 bundles.

 \setcounter{equation}{0}
 \section{Hard Orbifolds, Supersymmetry, and Characters}
 \label{repslong}
 Each distinct representation $R$, with real dimension
 $n \leq 7$, of any finite group $\Gamma\subset {\rm Aut}(T^n)$,
 can be used to define an orbifold $T^n/\Gamma$, and a
 corresponding compactification scheme in {\it M}-theory.
 In this section we describe some useful tools for efficiently
 accounting for large numbers of such constructions, and
 explain how these feed naturally into an algorithm for selecting
 those which satisfy a particular criterion: that the eleven-dimensional
 supercharge $Q$ have nonvanishing components at all points in
 $M^{11}$.
 This is done in two steps.  First we review some standard
 results pertaining to representations of finite groups.
 Then we explain some original technology which adapts these
 results to the special purpose of sifting through all possible representations and
 finding those which satisfy our criterion.

 \subsection{Representations of Finite Abelian Groups}
 \label{reps}
 Let $\Gamma$ be an abelian (commutative) group, and $\rho : \Gamma
 \rightarrow GL_n({\mathbb{C}})$ an $n$-dimensional complex matrix
 representation of $\Gamma$, i.e., $\rho$ is a homomophism of
 groups.  Since $\Gamma$ is abelian, $g_1\,g_2 = g_2\,g_1$ for all
 $g_i\in\Gamma$, and each element $g\in\Gamma$ equals its own
 conjugacy class.  A basic result in the theory of representations
 of finite groups states that for a group of order $q$, with $s$
 conjugacy classes, there are, up to equivalence, $s$ distinct
 irreducible representations
 $R_1, \ldots, R_s$ over $\C$.
 Moreover, if $R_i$ has dimension $n_i$, then
 \brr q := \sum_{i=1}^s(\,n_i\,)^2 \,.
 \err
 \cite[Theorem 2.3]{Leder}.  When applied to an abelian group
 $\Gamma$, this shows that each $n_i = 1$, i.e., that each of the
 irreducible representations of $\Gamma$ is itself a {\it
 character} (one dimensional representation) of the group.

 In fact, it is a simple matter to describe all the characters of a
 finite abelian group.  Any finite abelian group $\Gamma$ can be
 written as a direct product of $m$ cyclic groups of orders $q_1,
 \ldots, q_m$ respectively, so that
 \brr q=|\Gamma | = q_1\,q_2 \cdots q_m \,.
 \err
 A typical element of $\Gamma$ can be represented by the $m$-tuple
 $\vec{a} := (a_1, a_2, \ldots, a_m)$\,, where $0 \leq a_i <
 q_i$, with composition of elements given by componentwise addition
 followed by reduction of the $i$th component to its least
 nonnegative remainder modulo $r_i$, $i = 1, \ldots, m$.  Then
 corresponding to each $m$-tuple
 $\vec{c} := [c_1, c_2, \ldots, c_m]$\,,
 where $0 \leq c_i < r_i$, there exists a character
 \brr \Gamma_{\vec{c}}(\vec{a}) := \exp \left(2 \pi \imath \sum_{i=1}^m
      \left(\frac{a_i c_i}{r_i} \right) \right)
 \err
 of $\Gamma$, and all $q$ characters arise in this way
 \cite[Theorem 2.4]{Leder}. The identity element of $\Gamma$ corresponds to the
 trivial character.

 The obvious correspondence between the characters $\vec{c}$ and
 elements $\vec{a}$ of $\Gamma$ is not canonical.  Even though they
 are each composed of $m$-tuples of integers modulo $q_i$, $i = 1,
 \ldots, m$, the isomorphism between these is only well-defined up
 to an automorphism of the group $\Gamma$.

 Any $n$ dimensional representation of an abelian group $\Gamma$
 can be written as a direct sum of $n$ of its characters
 $\Gamma_{\vec{c}}$, i.e., by the data of an $n$-tuple
 $\{\vec{c}^1, \ldots, \vec{c}^n \}$ of $m$-tuples
 $\vec{c}^j :=[c_1^j, \ldots, c_m^j]$\,, for $j = 1, \ldots, n$.
 Two such representations are considered {\it equivalent} if these
 $n$-tuples agree as unordered lists.

 \subsection{Character Tables and $C$-Matrices}
 \label{cmatrices}
 The set of hard orbifolds $T^n/\Gamma$ is equivalent to the set of
 distinct $n$-dimensional representations of $\Gamma$
 consistent with the lattice that defines $T^n$.
 If $\Gamma$ is a finite abelian group, then
 these, in turn, are equivalent to the possible ways to
 order sets with elements chosen freely from among the
 characters of $\Gamma$, allowing for repetition.
 It is, therefore, a straightforward exercise, in principle, to
 construct
 comprehensive lists of hard orbifolds $T^n/\Gamma$.  This is so because it is
 also straightforward to determine the characters for any
 finite abelian group, using the following simple algorithm.
 (As described above, we shall limit our discussion to the case
 of pseudo-planar groups.)

 Each pseudo-planar group is given by the direct product of
 some number each of $\Z_2$, $\Z_3$ and $\Z_4$ factors.
 For each of these three ``elemental" groups, the list of
 characters are contained in the
 character tables exhibited in Table \ref{char234}.
 \begin{table}
 \begin{center}
 \begin{tabular}{c|cc}
 $\Z_2$ & $\Omega$ & $\Gamma$ \\[.1in]
 \hline
 &&\\[-.1in]
 {\bf 1} & + & + \\[.1in]
 $\a$ & + & $-$
 \end{tabular}
 \hspace{.3in}
 \begin{tabular}{c|ccc}
 $\Z_3$ & $\Omega$ & $\Sigma$ & $\bar{\Sigma}$ \\[.1in]
 \hline
 &&&\\[-.1in]
 {\bf 1} & + & + & + \\[.1in]
 $\b$ & + & $1/3$ & $-1/3$ \\[.1in]
 $\b^2$ & + & $-1/3$ & $1/3$
 \end{tabular}
 \hspace{.3in}
 \begin{tabular}{c|cccc}
 $\Z_4$ & $\Omega$ & $\Psi$ & $\Sigma$ & $\bar{\Psi}$ \\[.1in]
 \hline
 &&&\\[-.1in]
 {\bf 1} & + & + & + & +\\[.1in]
 $\g$ & + & $1/4$ & $1/2$ & $-1/4$ \\[.1in]
 $\g^2$ & + & $1/2$ & $+$ & $1/2$ \\[.1in]
 $\g^3$ & + & $-1/4$ & $1/2$ & $1/4$
 \end{tabular} \\[.1in]
 \caption{The ``elemental" character tables for the groups $\Z_2$,
 $\Z_3$ and $\Z_4$.}
 \label{char234}
 \end{center}
 \end{table}
 In these tables, the elements of the group are enumerated
 row-wise, while each column corresponds to a distinct character
 \footnote{Our convention differs from that used in the mathematical
 literature, wherein character tables typically list group elements as columns
 and characters as rows.  Our choice of convention is more suited to
 the particular application to physics described in this paper.}.
 For the case of $\Z_2$ the
 characters are real, i.e. each describes a group action on
 one real coordinate; in our case this corresponds to an action on
 an $A_1$ lattice; a plus sign in the table indicates a trivial
 action, while a minus sign indicates a sign change $x\to -x$ on the
 associated coordinate.
 For the groups $\Z_3$ and $\Z_4$ the nontrivial characters are complex;
 i.e. each describes a group action on a pair of real
 coordinates; in our case this corresponds to an action on
 an $A_2$ lattice; a plus sign indicates a trivial action,
 other rational numbers indicate the fraction of a complete
 counter-clockwise rotation in the plane spanned by the relevant
 $A_2$ lattice.  Such entries are defined modulo 1.

 It is useful to assemble the entries of a given character table for a
 group $\Gamma$ into a ``character matrix'' $\sigma(\Gamma)$.
 The data in Table \ref{char234} can be written as
 \brr \s(\Z_2)=\ba{cc}0& 0 \\0& 1/2 \ea
      \hspace{.1in}
      \s(\Z_3)=\ba{ccc}0&0&0\\0&1/3&$-1/3$\\0&$-1/3$&1/3\ea
      \hspace{.1in}
      \s(\Z_4)=\ba{cccc}0&0&0&0\\0&1/4&1/2&$-1/4$\\0&1/2&0&1/2\\0&$-1/4$&1/2&1/4\ea
      \,.
      \nonumber
 \err
 We adopt the convention that trivial actions (plus
 signs in the character tables) are represented in the
 character matrix with zeros, and
 parity reversals (minus signs in the character tables) are
 represented with the fraction 1/2.
 The character matrix for a generic pseudo-planar group with $m$
 elemental factors $\Gamma=G_1\times...\times G_m$,
 is obtained by combining the character matrices $\s(G_i)$
 as an outer sum. For instance, the character matrix
 $\s(\Z_2\times \Z_3)$ can be written as a a two-by-two array
 of three-by-three block matrices,
 wherein the upper left block is computed by adding the upper left
 entry in $\s(\Z_2)$ to the entire matrix $\s(\Z_3)$, the second
 block in the first row of blocks in $\s(\Z_2\times \Z_3)$
 is given by adding the entry $\s(\Z_2)_{12}$ to the entire
 matrix $\s(\Z_3)$, and so forth.   The matrix
 $\s(\Z_2\times\Z_3)$ formed in this way can be usefully re-expressed in
 terms of a character table, with the result
 shown in Table \ref{charz2z3}.  In Table \ref{charz2z3}, all rational entries
 are defined modulo 1.  Furthermore, a trivial action, denoted by a zero in the
 corresponding character matrix, is represented in the character table by a plus sign.
 Finally, on complex characters an entry 1/2, describing
 a 180 degree rotation, is represented in the table
 by a minus sign.
 \begin{table}
 \begin{center}
 \begin{tabular}{c|ccc|ccc}
 &&&\multicolumn{1}{c}{}&&&\\[-.1in]
 $\Z_2\times \Z_3$ & $\Omega$ & $\Sigma$ & \multicolumn{1}{c}{$\bar{\Sigma}$}
 & $\Lambda$ & $\Psi$ & $\bar{\Psi}$ \\[.1in]
 \hline
 &&&&&& \\[-.1in]
 $1$ & + & + & + & + & + & + \\[.1in]
 $\b$ & + & 1/3 & $-1/3$ & + & 1/3 & $-1/3$ \\[.1in]
 $\b^2$ & + & $-1/3$ & 1/3 & + & $-1/3$ & 1/3 \\[.1in]
 \cline{2-7}
 &&&&&&\\[-.1in]
 $\a$ & + & + & + & $-$ & $-$ & $-$ \\[.1in]
 $\a\b$ & + & 1/3 & $-1/3$ & $-$ & $-1/6$ & $1/6$  \\[.1in]
 $\a\b^2$ & + & $-1/3$ & 1/3 & $-$ & 1/6 & $-1/6$ \\[.1in]
 \end{tabular} \\[.2in]
 \caption{The character table for the group $\Z_2\times\Z_3$.}
 \label{charz2z3}
 \end{center}
 \end{table}
 Upon reconstituting the character matrix $\s(\Z_2\times\Z_3)$ into Table
 \ref{charz2z3} we have inserted a useful naming convention
 for the group elements and characters; we have named the
 order-two generating element $\a$ and the order-three generating
 element $\b$.  Similarly, we
 have named the trivial character $\Omega$, the order two
 character $\Lambda$, the order-three characters $\Sigma$ and
 $\bar{\Sigma}$ and the order-six characters $\Psi$ and
 $\bar{\Psi}$.

 By repeating the operation of combining character matrices as
 outer sums, in the manner described above, the character matrix
 and, equivalently, the character table for any pseudo-planar
 group can be generated readily from the three elemental character
 matrices $\s(\Z_2)$, $\s(\Z_3)$ and $\s(\Z_4)$.
 For an illustration in the case $\Gamma = ({\mathbb{Z}}_2)^3$, see
 Table \ref{charz2cube}.  Here $\Omega := \Gamma_{[0,0,0]}$ is the
 trivial character, and the group elements down the left-hand-side
 are indexed in the usual binary ordering:
 $$1 := (0,0,0) \, , \, \gamma := (0,0,1) \, , \, \beta := (0,1,0)
 \, , \, \ldots \, , \, \alpha \beta \gamma := (1,1,1) \ .$$

 In general, however, there is quite a lot of physically irrelevant redundancy
 in the full character table for a given orbifolding group $\Gamma$.
 For instance, in the case of $\Gamma=\Z_2\times\Z_3$,
 the characters which we have named $\Sigma$ and $\Psi$
 describe group actions on a complex coordinate $z$.
 However, if we describe these same characters
 in terms of their actions on the complex conjugate $\bar{z}$,
 these same characters would appear to act precisely as do
 $\bar{\Sigma}$ and $\bar{\Psi}$ on the original coordinate $z$.
 Thus, complex characters are physically indistinguishable from their conjugates.
 Since conjugate pairs of characters can be can be mapped
 into each other by a merely semantical renaming of the coordinates,
 we can more efficiently describe the relevant representation theory
 of this group by considering a restricted set of essential nontrivial
 characters.  For the case of $\Gamma=\Z_2\times \Z_3$, these
 would be $\Lambda$, $\Sigma$ and $\Psi$.  At the same time,
 elements with order greater than two have non-trivial inverses.
 These inverse elements have precisely the same locus of fixed-points
 in the physical space $T^n/\Gamma$ as do the original elements.
 So we can characterize the geometry of a given orbifold in terms
 of a representative set of essential non-trivial elements,
 thereby removing this second, physical, redundancy.

 The number of nontrivial representative elements of any finite abelian
 group is equivalent to the number of
 essential nontrivial characters. This number provides a ``physical rank"
 $r$ of the group.
 By including only the essential nontrivial elements and characters,
 we can replace the full character table with an ``abbreviated character table".
 For the case of $\Gamma=\Z_2\times\Z_3$, we would thereby replace Table \ref{charz2z3}
 with the abbreviated character table shown in Table \ref{abbrevz2z3}.
 \begin{table}
 \begin{center}
 \begin{tabular}{c|ccc}
 &&&\\[-.1in]
 $\Z_2\times \Z_3$ & $\Gamma$ & $\Sigma$ & $\Psi$ \\[.1in]
 \hline
 &&& \\[-.1in]
 $\a$ & 1/2 & 0 & 1/2  \\[.1in]
 $\b$ & 0 & 1/3 & $1/3$ \\[.1in]
 $\a\b$ & 1/2 & 1/3 & $-1/6$ \\[.1in]
 \end{tabular} \\[.2in]
 \caption{The abbreviated character table for the group $\Z_2\times\Z_3$.}
 \label{abbrevz2z3}
 \end{center}
 \end{table}
 Notice that we may choose at will the ordering of the elements (rows) and,
 independently, the ordering of the characters (columns) when we
 construct a character table; there is no a-priori canonical ordering.
 Notice, as well, that no information is sacrificed by
 replacing a full character table with an abbreviated character
 table.

 By multiplying the abbreviated character table by the order of the group
 we define an integer-valued, square $r\times r$
 matrix, which we denote $C(\Gamma)$.  For the
 case of $\Gamma=\Z_2\times\Z_3$, this is easily obtained from
 Table \ref{abbrevz2z3} by multiplying the entries by
 six, which is the order of $\Z_2\times\Z_3$.  In this way we
 determine
 \brr C(\Z_2\times\Z_3)=\ba{ccc} 3&0&3\\0&2&2\\3&2&-1\ea \,.
 \label{cmat23}\err
 Of course, owing to the freedom to independently rearrange the rows and the
 columns of the abbreviated character matrix, the corresponding matrix
 $C(\Gamma)$ is defined only up to similar
 reorderings.  However, such flexibility can always be used
 to render the $C$-matrix symmetrical.  We deem this canonical.
 It is also possible, while keeping $C(\Gamma)$
 symmetrical, to arrange the rows and columns so that the
 corresponding elements and characters have
 monotonically increasing order.  This too, we deem canonical.

 All the information described by Table \ref{charz2z3} is also
 contained in the matrix $C(\Z_2\times\Z_3)$ shown in
 (\ref{cmat23}).
 It is interesting that quite a lot of information pertaining to
 properties of any finite abelian group, including the complete
 representation theory can be codified in a symmetric matrix
 $C\in GL(\,r\,,\,\Z\,)$.  As it turns out, the matrices
 $C(\Gamma)$ are valuable tools in the search for supersymmetric
 orbifolds.  The matrices $C(\Gamma)$ for each of the
 pseudo-planar groups with group order $\le 12$ are listed in
 Appendix \ref{cmatlist}.

 \subsection{The Enumeration of Distinct Orbifolds}
 \label{orbenum}

 A representation of $\Gamma$ is designated
 by choosing a set of real and complex characters, including the
 possibility of degeneracy, from the list
 of essential nontrivial characters.  Generally,
 order-two characters are real, while characters with higher order
 are complex.  Therefore, if we select $a$ order-two characters and
 $b$ higher-order characters, the corresponding representation
 will act on $n=a+2\,b$ real dimensions, $2\,b$ of which are
 complexified.  Since the set of essential nontrivial characters
 correlates with the columns of the matrix $C(\Gamma)$, we can
 unambiguously designate a representation by an ordered list of
 $r$ multiplicities, each indicating the number of real coordinates
 transforming according to a corresponding character.
 The ordering of the multiplicities
 corresponds to the ordering of the characters described by the
 rows of the $C$-matrix.  Of course, the multiplicities
 corresponding to complex characters are necessarily even, while
 those corresponding to order-two characters may be even or odd.

 As an example, in the case of the group $\Gamma=\Z_2\times \Z_3$,
 the physical rank is 3, and the corresponding $C$-matrix is
 given by (\ref{cmat23}).  In this case, each representation is
 given by a $3$-tuple, $R=(\,a_1\,,\,a_2\,,\,a_3\,)$,
 where $a_1$, $a_2$ and $a_3$ are the number of real coordinates
 transforming according to the characters $\Lambda$, $\Sigma$ and
 $\Psi$, respectively.  In this case, $a_1\in {\mathbb{N}}$ since the
 character $\Gamma$ is real, and $a_{2,3}\in 2\,{\mathbb{N}}$
 since the characters $\Sigma$ and $\Psi$ are complex.  To be
 quite specific, the representation of $\Z_2\times\Z_3$ described by
 the three-tuple $R=(\,3\,2\,2\,)$ is a representation
 which acts on $3+2+2=7$ real coordinates.  In this case, however,
 four of the real coordinates are complexified as two complex coordinates.
 The first multiplicity (3) in the label (322) indicates that three real
 coordinates, say $x_{1,2,3}$ transform according to the character
 $\Lambda$, the second multiplicity (2) indicates that
 two real coordinates, combined into one complex
 coordinate, say $z_1$, transform according to the character $\Sigma$,
 and the third multiplicity (2)  indicates that one more complex
 coordinate, say $z_2$, transforms according to the character
 $\Psi$.  The corresponding group actions are shown
 in Table \ref{rep322}.
 \begin{table}
 \begin{center}
 \begin{tabular}{c|ccccc}
 $\Z_2\times\Z_3$ & $x_1$ & $x_2$ & $x_3$ & $z_1$ & $z_2$ \\[.1in]
 \hline
 &&&&\\[-.1in]
 $\a$ & $-$ & $-$ & $-$ & + & $-$ \\[.1in]
 $\b$ & + & + & + & 1/3 & 1/3 \\[.1in]
 $\a\b$ & $-$ & $-$ & $-$ & 1/3 & $-1/6$
 \end{tabular} \\[.1in]
 \caption{The representation $R=(322)$ of the group
 $\Z_2\times\Z_3$}
 \label{rep322}
 \end{center}
 \end{table}

 The particular orbifold $T^n/\Gamma$ which corresponds to a given
 representation $R$ generically includes a locus of special points which
 remain invariant under elements of $\Gamma$.  These
 generically constitute hyperplanes of various dimensionalities
 which intersect, forming an intricate network.  One of our
 primary concerns is to decide what sorts of physics,
 in the form of localized states, are described by these planes
 and their intersections.  In the next section we will describe in
 detail how one studies the issue of how many supercharges are
 retained on these.  A primary consideration in this regard
 is, of course, to describe the number of dimensions which are
 spanned by the fixed-planes associated with each group element.

 There is a simple formula which allows one to
 compute the set of dimensions corresponding to
 the $r$ representative nontrivial elements.  This formula is most easily
 described in terms of another useful matrix, which we call $M(\Gamma)$,
 obtained from $C(\Gamma)$ by replacing all zero
 entries with ones, and all nonzero entries by zeros.  By way of
 illustration, we focus again on the example
 $\Gamma=\Z_2\times\Z_3$.  In this case, this prescription,
 applied to $C(\Z_2\times\Z_3)$, as given in (\ref{cmat23}), yields
 \brr M(\Z_2\times\Z_3)=\ba{ccc} 0 & 1 & 0 \\
      1 & 0 & 0 \\
      0 & 0 & 0 \ea \,.
 \err
 For a given orbifold, described by the representation $R$ of
 a group $\Gamma$, the dimensionality of the fixed-planes
 of each representative element is described by another
 $r$-tuple, $d(R)$, given by
 \brr d(R)=(\,11-n\,)\,{\bf 1}+R\,M \,,
 \err
 where ${\bf 1}$ is the row vector with ones in each entry
 and $n=R\,\cdot\,{\bf 1}=\sum_i a_i$.
 As an example, for the particular orbifold described by $R=(322)$,
 we compute $n=7$ and
 \brr d(122) &=& (11-7)\,\ba{ccc}1&1&1\ea
      +\ba{ccc}3&2&2\ea\,\ba{ccc}0&1&0\\1&0&0\\0&0&0\ea
      \nonumber\\[.1in]
      &=& \ba{ccc}6&7&4\ea \,.
 \err
 Thus, the respective fixed-planes associated with the elements $\a$, $\b$ and $\a\b$
 have dimensionality 6, 7 and 4.  This result can be verified from the
 precise group actions in Table \ref{rep322}.

 Now we have all the information we need to form comprehensive
 lists of all pseudo-planar orbifolds, including all the data pertaining to the
 group actions.  First we choose a group $\Gamma$.   Then,
 we form lists of orbifolds $T^n/\Gamma$, for
 each value of $1\le n\le 7$ by sequencing through the ordered partitions
 of $n$ into $r$ nonnegative integers, in the manner described above.
 For the case of $T^7/(\Z_2\times\Z_3)$ orbifolds, for example, we create the
 the sequence of 3-tuples $(\,a_1\,,\,a_2\,,\,a_3\,)$ which describe ordered partitions
 of $n=7$ into sums of $r=3$ nonnegative integers, subject to the constraints that
 $a_1\in\N$ and $a_{2,3}\in 2\,\N$.  The complete list of such $3$-tuples
 is given in the usual ascending order in mod 7 arithmetic, as
 (106), (124), (142), (160), (304), (322), (340), (502), (520), (700).
 We describe these ordered sets of multiplicities as
 orbifold {\it labels}.
 In each case the corresponding group actions can be determined
 by dividing the $C$-matrix by the group order and then selecting rows from
 this divided $C$-matrix with the appropriate multiplicity
 indicated by the corresponding label.
 The group actions for the orbifold $T^7/(\Z_2\times\Z_3)_{(322)}$,
 described above, were obtained in precisely this way, and
 stand as an example of this methodology.

 \subsection{The Periodic Table}
 We use the algorithm described in the previous paragraph to
 systematically cycle through each orbifold label of each
 pseudo-planar group, obtaining all of the relevant group actions
 in each case.  In each instance, each element of $\Gamma$ can
 be represented, by arranging the coordinates judiciously, by a
 set of three fractional rotations $(\,f_1\,,\,f_2\,,\,f_3\,)$
 describing counterclockwise rotations in respective planes spanned by three
 $A_2$ lattices, plus possibility of a parity reversal in one real
 coordinate, which we codify as the binary choice $P\in \{0,1\}$,
 describing the respective absence or presence of a parity reversal.

 As it turns out, the values of $(\,f_1\,,\,f_2\,,\,f_3\,|\,P\,)$ for
 each element of $\Gamma$ corresponding to a given orbifold
 label provide all the data necessary to resolve the amount of
 supersymmetry on the corresponding orbifold plane.  The precise
 corrrespondence is derived in the next section, where it is
 presented as a supersymmetric restriction theorem.  This result
 says that an orbifold plane is supersymmetric if and only if
 there is at least one way to add or subtract the three corresponding
 fractions $f_i$ to obtain, in the case $P=0$, an even integer,
 or, in the case $P=1$, any integer (even or odd).

 For our restricted class of lattices, a given element
 is compatible only if the three $f_i$ are each elements of the set
 $\{\,0\,,\,1/2\,,\,\pm 1/3\,,\,\pm 1/4\,,\,\pm 1/6\,\}$ mod 1.
 (There are, therefore $7^3\times 2=686$ possibilities for each
 element.) It is possible to have compatible elements which do
 not have compatible products.
 For example, in the case $\Gamma=\Z_3\times\Z_4$, there are several models
 which pass the criteria of the supersymmetric restriction theorem, except
 that the representations in question involve order-twelve rotations
 in at least one plane.  These would be acceptable as
 supersymmetric orbifolds of $\R^n$, but not of $T^n$, because
 there is no lattice in $\C$ compatible with such a rotation.
 The number of global orbifolds $T^n/\Gamma$ is therefore
 much smaller than the number of orbifold singularities
 which can be modelled locally as $\R^n/\Gamma$.

 The search for supersymmetric orbifolds consists of four
 steps.  First, for a given choice of $\Gamma$ and $n$,
 we generate the complete list of compatible orbifold labels.
 Second, for each orbifold label, we use the matrix $C(\Gamma)$
 to determine the data $(\,f_1\,,\,f_2\,,\,f_3\,|\,P\,)$ for each
 of the $r$ repressentative elements.  Third, we apply the
 supersymmetric restriction theorem to remove each orbifold which
 has any element whose data does not meet the restriction
 criterion.  Fourth, we examine the list of orbifolds which
 satisfy these restrictions, and we remove cases which are
 redundant.  We have created a number of Mathematica
 functions which fully automate this process, and have used this
 to generate Table \ref{scan2} which appears in the introduction.
 In this way, we can easily generalize our periodic table to
 arbitrary group order.

 Our periodic table includes all of the hard global {\it M}-theory
 orbifolds described previously by other authors as well as by
 ourselves.  For instance the $S^1/\Z_2$ model corresponds to
 the original {\it M}-theory model described in \cite{hw1, hw2}.
 The four $T^4/\Gamma$ models correspond to the four global
 orbifold limits of $K3$. The $T^5/\Z_2$ model was discussed
 in \cite{dasmuk, witt5}.  (It was the study of that simple model
 which first implicated wandering five-branes
 as a means of unifying ostensibly unique vacua into classes
 linked by phase transitions.) The four starred models $T^5/\Gamma$
 correspond to the four global orbifold limits of
 $K3\times S^1/\Z_2$ and were studied in
 \cite{mlo, phase, hetk3, ksty}.  The $T^7/(\Z_2)^3$ model
 with label (2220001) was described in \cite{DFO} and the
 $T^7/(\Z_2\times\Z_2\times\Z_3)$ with label (0122022) was
 described in \cite{chiralz}.  Finally, resolutions of the
 softenings of the $T^7/(\Z_2)^3$ model
 with label (1111111) were presented by Joyce in \cite{JoyJDG1, JoyJDG2, Joyce}
 as prototype $G_2$ manifolds, and studied by Acharya as candidate
 $M$-theory compactification spaces \cite{AchA, AchB} (see also
 Section \ref{softsect} below).

 \setcounter{equation}{0}
 \section{A Supersymmetric Restriction Theorem}
 \label{theorem}
 In the bulk of $M^{11}$ (i.e., at all points not within the
 locus of orbifold fixed-planes) the eleven-dimensional supercharge is
 completely preserved.  It is within the locus of fixed-planes, therefore,
 that the issue of supercharge preservation becomes important.
 Since the supercharge transforms nontrivially under elements of
 $\Gamma$, only those components of $Q$ which remain invariant are
 not projected to zero on those spacetime points inert under those
 same elements.  The invariant components of $Q$ typically resolve as a
 $d$ dimensional spinor, where $d$ is the total dimension of the
 invariant locus associated with that element near a given point.
 Precisely how many irreducible $SO(d-1,1)$-spinors are included in
 this set determines the amount of local sypersymmetry preserved on
 that fixed-plane.  For the sorts of pseudo-planar orbifolds
 defined above, it is possible to delineate a concise criterion
 for selecting those which, in this way, retain
 supersymmetry.
 In this section, which is relatively technical,
 we derive this supersymmetric restriction theorem.  We
 start by establishing notational conventions and stating the result,
 and then prove it by analyzing the question of how the spinorial supercharge is
 influenced by lifts of various elements
 of finite subgroups of $SO(10,1)$.

\subsection{Notations}
\label{susynotat}

 We use spacetime coordinates $x^I\equiv \{\,x^0\,,\,x^i\,\}$,
 where $i = 1,..., 3,5,...11$, and
 Gamma matrices $\Gamma_I$, which satisfy the Clifford algebra
 $\{\Gamma_I, \Gamma_J \} = 2\,\eta_{IJ}$,
 where $\eta_{IJ}=\mbox{diag}(-+ \cdots +)$ is the flat
 metric.  The Gamma matrices are chosen such that $\G_0$ is
 anti-Hermitian, $\G_0^\dag = - \G_0$, and
 $\G_i$ are Hermitian, $\G_i^\dag = \G_i$.  It is sometimes
 useful to define $\Gamma_4=i\,\Gamma_0$.
 Another useful identity is $\G_{11} = i\,\G_1 \cdots \G_{10}$.
 The matrices $\G_{IJ} = \ft12\,[\,\G_I\,,\,\G_J\,]$ are the
 generators of $\mbox{spin}(11)$.
 We define a complex structure by writing the six real coordinates
 $x_{5,...,10}$ in terms of three complex coordinates,
 according to $z_{1,2,3}\equiv x_{5,7,9}+i\,x_{6,8,10}$.

 Consider an element which acts as simultaneous rotations in three
 complex planes, with coordinates $z_1$, $z_2$, and $z_3$, and
 possibly a parity flip on one real coordinate $x^{11}$,
 \brr \a : (\,z_1\,,\,z_2\,,\,z_3\,;\,x^{11}\,) \to
      (\,e^{i\,\q_1}\,z_1\,,\,e^{i\,\q_2}\,z_2\,,
      \,e^{i\,\q_3}\,z_3\,;\,(-)^P x^{11}\,) \,,
 \label{alphaeq} \err
 with $\q_i = 2\,\pi\,f_i$.  The three rational
 numbers $f_i$ describe the fraction of a complete rotation
 imparted respectively on the three complex planes.  The parameter
 $P \in \{0,1\}$ indicates whether or not the element includes a
 parity flip.  The order of $\alpha$ is given by
 the least common positive integer multiple of the denominators of
 the reduced form of the three $f_i$'s and also $(1/2)^P$.

 For a given choice of
 $(\,f_1\,,\,f_2\,,\,f_3\,|\,P\,)$, equation (\ref{alphaeq}) describes a
 particular global $SO(10,1)$ transformation
 plus the possibility of a parity transformation.  On spinors this
 induces $\psi \to \Omega\,\psi$, where
 \brr \Omega = \exp\,(\,\ft14\,\q^{IJ}\,\G_{IJ}\,)\,(\,\G_{11}\,)^P
 \label{Omegaeq}\err
 and where $\q^{IJ}$ are the parameters of the $SO(10,1)$ transformation.  From
 (\ref{alphaeq}) we read off the non-vanishing parameters as
 $\q^{5,6}=- \q^{6,5} \equiv \q_1$, $\q^{7,8}=- \q^{8,7} \equiv
 \q_2$, and $\q^{9,10}=- \q^{10,9} \equiv \q_3$.  Thus we can
 rewrite (\ref{Omegaeq}) as
 \brr \Omega = \exp\,(\,\ft12\,\theta_1\,\Gamma_{5,6}\,)\,
      \exp\,(\,\ft12\,\theta_2\,\Gamma_{7,8}\,)\,
      \exp\,(\,\ft12\,\theta_3\,\Gamma_{9,10}\,)(\,\G_{11}\,)^P \,.
 \label{Omeganew} \err
 The supercharge $Q$ is an eleven-dimensional Majorana spinor,
 which transforms precisely as
 $Q\to \Omega\,Q$.  It is useful to append
 subscripts to spinors, and to spinorial operators, to indicate
 dimensionality.  We
 thus write the eleven-dimensional supercharge $Q$ as $Q_{(11)}$ to
 indicate that this field takes its values in $\mbox{spin}(11)$.
 Similarly, we write the operator $\Omega$, defined in (\ref{Omeganew}), as
 $\Omega_{(11)}$ to indicate that this object operates on
 eleven-dimensional spinors.

 If the fixed-point locus associated with an element
 (\ref{alphaeq}) has dimensionality $d$, then, in the neighborhood
 of this locus, the structure group is broken
 from $SO(10,1)$ down to $SO(d-1,1)\times SO(11-d)$.
 Accordingly, we write the supercharge as a tensor product of
 an $SO(d-1,1)$ fixed-plane spinor and an $SO(11-d)$ ``normal" spinor,
 as $Q_{(11)}=Q_{(d)}\otimes Q_{(11-d)}$.  Similarly, the
 operator $\Omega$ decomposes as
 $\Omega_{(11)}=\Omega_{(d)}\otimes \Omega_{(11-d)}$.
 In order to resolve the amount of unbroken fixed-plane supersymmetry, we
 solve the equation $Q_{(11)}=\Omega_{(11)}\,Q_{(11)}$,
 and then count the degrees of freedom which describe the most general solution.

 The analysis described below is completely general, modulo irrelevant
 reordering of coordinates.  Thus, in this section we consider the element
 $(\,0\,,\,0\,,\,f\,|\,0\,)$ equivalent to $(\,0\,,\,f\,,\,0\,|\,0\,)$,
 for instance.

 \subsection{Statement of the Theorem}
 \label{susythm}

 The conclusions which we draw in each of the
 seven cases discussed in Section \ref{susypf} below can be easily summarized as follows.
 The condition for supersymmetry on a fixed-plane
 associated with any element $\a\in \Gamma$, of the special
 sort characterized by (\ref{alphaeq}), is
 \brr f_1\pm f_2\pm f_3\in \left\{ \begin{array}{c}
      2\,\Z \hspace{.4in};\,\, P=0 \\[.2in]
      \Z \hspace{.4in};\,\, P=1
      \end{array}\right.
 \label{cod}\err
 for any one of the four possible choices of unspecified
 signs.   If condition (\ref{cod}) is satisfied, then
 supersymmetry is generically reduced to the minimal amount
 possible in the dimensionality
 of the fixed-plane.  Exceptions occur in cases where $P=1$, when
 one of the sums in (\ref{cod}) gives an even integer and another gives an
 odd integer.  In such cases, the fixed-plane supersymmetry is
 twice the minimal amount.
 These results are reflected in Table \ref{table1}.
 \begin{table}[t]
 \begin{center}
 \begin{tabular}{ccc}
 $(\,f_1\,,\,f_2\,,\,f_3\,|\,(-)^P\,)$ & $d_{\rm fixed}$ & SUSY \\[.1in]
 \hline
 $(\,0\,,\,0\,,\,0\,|\,-\,)$ & 10 & $1/2$ \\[.1in]
 $(\,0\,,\,0,f\,|\,+\,)$ & 9 & NONE \\[.1in]
 $(\,0\,,\,0\,,\,f\,|\,-\,)$ & 8 & NONE \\[.1in]
 $(\,0\,,\,f\,,\,\pm f\,|\,+\,)$ & 7 & $1/2$ \\[.1in]
 $(\,0\,,\,f\,,\,\pm f\,|\,-\,)$ & 6 & $1/4^*$ \\[.1in]
 $(\,f_1\,,\,f_2\,,\,f_3\,|\,+\,)$ & 5 & $1/4$ \\[.1in]
 $(\,f_1\,,\,f_2\,,\,f_3\,|\,-\,)$ & 4 & $1/8^*$ \\[.1in]
 \end{tabular}
 \end{center}
 \caption{The seven sorts of elements,
 listed along with the associated fixed-plane dimensionality
 and the generic amount of supersymmetry retained when the
 conditions listed in (\ref{cod}) are satisfied.
 The $d=6$ and the $d=4$ cases have exceptions.
 In the $d=6$ case, if $f = 1/2$ then the supersymmetry is
 merely halved, not quartered.  In the $d=4$ case
 if $|f_1|$, $|f_2|$ and $|f_3|$ are drawn, one each, from
 either set $(1/2,1/3,1/6)$ or $(1/2,1/4,1/4)$ then supersymmetry is
 merely quartered, not eighthed. }
 \label{table1}
 \end{table}

 Note that this result applies to orbifolds
 $\R^7/\Gamma$ as well as to orbifolds $T^7/\Gamma$.
 In the former case, there is no restriction on the
 choices of the $f_i$'s other than that they should be
 rational.  In the latter case, there are additional
 constraints which follow from the requirement that
 $\Gamma\subset {\rm Aut}(\,T^7\,)$.  For the cases
 described in Section \ref{susypf}, this requirement
 amounts to the restriction that $f_{1,2,3}$ must be chosen
 from the set $\{\,0\,,\,1/2\,,\,\pm 1/3\,,\,\pm 1/4\,,\,\pm 1/6\,\}$
 mod 1.

\subsection{Proof of the Theorem}
\label{susypf}

The result is established by explicitly analyzing how the
spinorial supercharge is influenced by lifts of various elements
 of finite subgroups of $SO(10,1)$.\\

\noindent
 {\it Ten-dimensional fixed-planes:}\\[.1in]
 The only transformation (\ref{alphaeq}) which has
 ten-dimensional fixed-planes is an order-two
 element acting as $(f_1,f_2,f_3\,|\,P)=(0,0,0\,|\,1\,)$.
 In this case, the supercharge transforms as
 $Q_{(11)} \to \Gamma_{11}\,Q_{(11)}$.
 We decompose $Q_{(11)}$  according to
 $Q_{(11)}=Q_{(10)R}\oplus Q_{(10)L}$, where
 $Q_{(10)R,L}=\pm\Gamma_{11}\,Q_{(10)R,L}$ are ten-dimensional
 Majorana-Weyl projections of $Q_{(11)}$. The fixed-plane condition,
 $Q_{(11)}=\Omega_{(11)}\,Q_{(11)}$ can now be written
 \brr Q_{(10)\,R,L}  =
      \pm Q_{(10)\,R,L} \,.
 \err
 This equation is easy to solve.  On the fixed-plane we have
 $Q_{(11)}\to Q_{(10)R}$.  Thus, the bulk supersymmetry is
 {\it halved} on the fixed-planes.\\

\noindent
 {\it Nine-dimensional fixed-planes:}\\[.1in]
 Nine-dimensional fixed-planes correspond to elements
 $(f_1,f_2,f_3\,|\,P)=(0,0,f\,|\,0\,)$,
 where $f\ne 0$ mod 1.
 In this case, the supercharge transforms as
 $Q_{(11)}\to \Omega_{(11)}\,Q_{(11)}$, where
 \brr \Omega_{(11)} =
      \exp\,(\,{\pi\,f\,\Gamma_9\Gamma_{10}}\,) \,.
 \label{to9}\err
 We write $Q_{(11)}$ as a tensor product of a nine-dimensional fixed-plane
 spinor with an $SO(2)$ ``normal" spinor, according to  $Q_{(11)}=Q_{(9)}\otimes Q_{(2)}$.
 Similarly, we use the following representation for the Gamma
 matrices,
 \brr \Gamma_\mu &=& \widehat{\Gamma}_\mu\otimes {\bf 1}
      \hspace{.3in};\,\,\mu=1,...,8
      \nonumber\\[.1in]
      \Gamma_{8+i} &=& \widehat{\Gamma}_9\otimes \sigma_i
      \hspace{.3in};\,\,i=1,2 \,,
      \nonumber\\[.1in]
      \Gamma_{11} &=& \widehat{\Gamma}_9\otimes \s_3 \,,
 \label{gam8}\err
 where $\{\,\widehat{\Gamma}_\mu\,,\,\widehat{\Gamma}_9\,\}$ are
 nine-dimensional Gamma matrices and $\s_i$ are
 the Pauli matrices.  In addition, we
 decompose the normal spinors according to
 $Q_{(2)}=Q_{(2)R}+ Q_{(2)L}$
 where $Q_{(2)R,L}=\pm\,\s_3\,Q_{(2)R,L}$.
 Using these conventions, and also the identity
 $\s_1\s_2=i\,\s_3$, we can write
 \brr Q_{(11)} &=& Q_{(9)}\otimes (\,Q_{(2)R}+ Q_{(2)L}\,)
      \nonumber\\[.1in]
      \Omega_{(11)} &=& \exp(\,i\,\pi\,f\,{\bf 1}\otimes \s_3\,)
      \,.
 \err
 The fixed-plane condition $Q_{(11)}=\Omega_{(11)}\,Q_{(11)}$
 can now be written as
 \brr Q_{(9)}\otimes Q_{(2)R,L}=e^{\pm i\,\pi\,f}\,Q_{(9)}\otimes Q_{(2)R,L} \,.
 \err
 Because there are no nontrivial solutions to either of these equations,
 subject to the restriction that $f\ne 0$ mod 1, we conclude
 that, as we approach a fixed-nine-plane, $Q_{(11)}\to 0$.
 Thus supersymmetry is broken completely on any nine-dimensional
 fixed-plane associated with an element (\ref{alphaeq}).\\

\noindent
 {\it Eight-dimensional fixed-planes:}\\[.1in]
 Eight-dimensional fixed-planes correspond to elements
 $(\,f_1\,,\,f_2\,,\,f_3\,|\,P\,)=(\,0\,,\,0\,,\,f\,|\,1\,)$,
 where $f\ne 0$ mod 1.
 In this case, the supercharge transforms as
 $Q_{(11)}\to\Omega_{(11)}\,Q_{(11)}$, where
 \brr \Omega_{(11)} =
      \exp\,(\,{\pi\,f\,\Gamma_9\,\Gamma_{10}}\,)\,\Gamma_{11} \,.
 \label{to8}\err
 We write $Q_{(11)}$ as a tensor product of an eight-dimensional
 fixed-plane spinor with an $SO(3)$ ``normal" spinor, according to
 $Q_{(11)}=Q_{(8)}\otimes Q_{(3)}$, and we represent the Gamma matrices
 precisely as in (\ref{gam8}).   In addition we decompose the
 $SO(7,1)$ spinors via $Q_{(8)}=Q_{(8)\,R}+ Q_{(8)\,L}$
 where $Q_{(8)R,L}=\pm\,\widehat{\Gamma}_9\,Q_{(8)R,L}$, and
 write the normal spinors as
 $Q_{(3)}=Q_{(2)\,R}+ Q_{(2)\,L}$,
 where $Q_{(2)\,R,L}=\pm\,\s_3\,Q_{(2)\,R,L}$.
 We introduce a useful shorthand notation whereby
 $L\equiv Q_{(2)\,L}$ and  $R \equiv Q_{(2)\,R}$.
 Using these conventions, and also the identity
 $\s_1\,\s_2=i\,\s_3$, we can write
 \brr Q_{(11)} &=& (\,Q_{(8)\,R}+Q_{(8)\,L}\,)\otimes (\,R + L\,)
      \nonumber\\[.1in]
      \Omega_{(11)} &=& \exp({i\,\pi\,f\,{\bf 1}\otimes \sigma_3})\,
      \widehat{\Gamma}_9\otimes \sigma_3 \,.
 \err
 The fixed-plane condition $Q_{(11)}=\Omega_{(11)}\,Q_{(11)}$ can
 now be written
 \brr Q_{(8)\,R,L}\otimes R &=& \pm e^{i\,\pi\,f}\,Q_{(8)\,R,L}\otimes R
      \nonumber\\[.1in]
      Q_{(8)\,R,L}\otimes L &=& \mp e^{-i\,\pi\,f}\,Q_{(8)\,R,L}\otimes L\,.
 \err
 Because there are no nontrivial solutions to any of these four equations,
 subject to the restriction that $f\ne 0$ mod 1, we conclude
 that, as we approach an eight-dimensional fixed-plane, $Q_{(11)}\to 0$.
 Thus, supersymmetry is broken completely on any eight-dimensional
 fixed-plane associated with an element (\ref{alphaeq}).\\

\noindent
 {\it Seven-dimensional fixed-planes:}\\[.1in]
 Seven-dimensional fixed-planes correspond to elements
 $(\,f_1\,,\,f_2\,,\,f_3\,|\,P\,)=(\,0\,,\,f\,,\,f'\,|\,0\,)$,
 where $f\ne 0$ mod 1 and $f'\ne 0$ mod 1.
 In this case, the supercharge transforms as
 $Q_{(11)}\to \Omega_{(11)}\,Q_{(11)}$, where
 \brr \Omega_{(11)} &=&
      \exp\,(\,\pi\,f\,\Gamma_7\,\Gamma_{8}\,)\,
      \exp\,(\,\pi\,f'\,\Gamma_9\,\Gamma_{10}\,)\,.
  \label{to7}\err
 We write $Q_{(11)}$ as a tensor product of a seven-dimensional
 fixed-plane spinor with an $SO(4)$ ``normal" spinor according to
 $Q_{(11)}=Q_{(7)}\otimes Q_{(4)}$.  Similarly, we use the following
 representation for the Gamma matrices,
 \brr \Gamma_\mu &=& \widehat{\Gamma}_\mu\otimes {\bf 1}
      \hspace{.3in};\,\,\mu=1,...,6
      \nonumber\\[.1in]
      \Gamma_{6+i} &=& \widehat{\Gamma}_7\otimes \gamma_i
      \hspace{.3in};\,\,i=1,...,4 \,,
      \nonumber\\[.1in]
      \Gamma_{11} &=& \widehat{\Gamma}_7\otimes \gamma_5 \,,
 \label{gam7}\err
 where $\{\,\widehat{\Gamma}_\mu\,,\,\widehat{\Gamma}_7\,\}$ are the seven
 dimensional Gamma matrices
 and $\gamma_i$ are four dimensional gamma matrices.
 In addition, we decompose the normal $SO(4)$ spinors
 using the decomposition $SO(4)\to SO(2)\times SO(2)$,
 where $SO(2)\times SO(2)$ is a convenient maximal subgroup of
 $SO(4)$.  Thus, we write the ``normal" spinor as $Q_{(4)}=Q_{(2)}\otimes Q_{(2)}$,
 and the four dimensional gamma matrices as
 $\g_{1,2}=\s_{1,2}\otimes {\bf 1}$ and
 $\g_{3,4}=\s_3\otimes \s_{1,2}$, where $\s_i$ are Pauli matrices.
 Also, using the identities $\g_5=\g_1\g_2\g_3\g_4$ and $\s_1\s_2=i\,\s_3$,
 we derive $\g_5=-\s_3\otimes \s_3$.
 This allows us to rewrite
 (\ref{gam7}) as
 \brr \Gamma_{1,...,6} &=& \widehat{\Gamma}_{1,...,6}\otimes
      {\bf 1}\otimes {\bf 1}
      \nonumber\\[.1in]
      \Gamma_{7,8} &=& \widehat{\Gamma}_7\otimes\s_{1,2}\otimes
      {\bf 1}
      \nonumber\\[.1in]
      \Gamma_{9,10} &=& \widehat{\Gamma}_7\otimes\s_3\otimes
      \s_{1,2}
      \nonumber\\[.1in]
      \Gamma_{11} &=& -\widehat{\Gamma}_7\otimes\s_3\otimes\s_3
      \,.
 \label{gam7y}\err
 Note that the normal spinor can be decomposed as
 $Q_{(4)}=Q_{(4)\,R}+Q_{(4)\,L}$ where
 $Q_{(4)\,R,L}=\pm\g_5\,Q_{(4)\,R,L}$.
 In terms of $SO(2)\times SO(2)\subset SO(4)$, however,
 these same objects can be written as
 \brr Q_{(4)R} &=& Q_{(2)R}\otimes Q_{(2)L}+Q_{(2)L}\otimes Q_{(2)R}
      \nonumber\\[.1in]
      Q_{(4)L} &=& Q_{(2)R}\otimes Q_{(2)R}+Q_{(2)L}\otimes Q_{(2)L}
      \,,
 \label{de4}\err
 where $Q_{(2)\,R,L}=\pm \s_3\,Q_{(2)\,R,L}$.
 Using these results we can decompose the eleven-dimensional
 supercharge in two steps. First, we write
 $Q_{(11)} = Q_{(7)} \otimes (\,Q_{(4)\,R}+Q_{(4)\,L}\,)$ and then
 we rewrite the terms $Q_{(4)\,R,L}$ using (\ref{de4}).
 We now define $LL\equiv Q_{(2)L}\otimes Q_{(2)L}$, and similar
 expressions for $LR$, $RL$, and $RR$.
 Using these definitions, and also the conventions introduced above, we can
 write
 \brr Q_{(11)} &=& Q_{(7)}\otimes \bpl\,
      LL + LR + RL + RR \,\bpr
      \nonumber\\[.1in]
      \Omega_{(11)} &=&
      \exp(\,i\,\pi\,f\,{\bf 1}\otimes \s_3\otimes {\bf 1}
      +i\,\pi\,f'\,{\bf 1}\otimes {\bf 1}\otimes \s_3\,) \,.
 \label{new22}\err
 The fixed-plane condition $Q_{(11)}=\Omega_{(11)}\,Q_{(11)}$ can now be
 written as
 \brr Q_{(7)}\otimes LL &=& e^{-i\,\pi\,(f+f')}\,Q_{(7)}\otimes LL
      \nonumber\\[.1in]
      Q_{(7)}\otimes LR &=& e^{-i\,\pi\,(f-f')}\,Q_{(7)}\otimes LR
      \nonumber\\[.1in]
      Q_{(7)}\otimes RL &=& e^{+i\,\pi\,(f-f')}\,Q_{(7)}\otimes RL
      \nonumber\\[.1in]
      Q_{(7)}\otimes RR &=& e^{+i\,\pi\,(f+f')}\,Q_{(7)}\otimes RR
      \,.
 \err
 Therefore, as we approach a fixed-seven-plane, the bulk supercharge is projected according to
 \brr Q_{(11)}\to \left\{\begin{array}{c}
      \,Q_{(7)}\otimes \bpl\, LL + RR \,
      \bpr \hspace{.4in};\,\, f+f' \in 2\,\Z \\[.1in]
      Q_{(7)}\otimes \bpl\, LR + RL \,
      \bpr \hspace{.4in};\,\, f-f' \in 2\,\Z
      \end{array}\right. \,,
 \label{q7b}\err
 and is projected to zero if neither of these conditions are met.
 Thus we retain some supersymmetry on a fixed-seven-plane
 if and only if one of the two sums $f\pm f'$ is an even integer.
 This is the case only if $f=\pm f'$ mod 1.
 In such cases, the bulk supersymmetry is {\it halved} on
 the fixed-plane in question. \\

\noindent
 {\it Six-dimensional fixed-planes}\\[.1in]
 Six-dimensional fixed-planes correspond to elements
 $(f_1,f_2,f_3\,|\,P)=(0,f,f'\,|\,1\,)$,
 where $f\ne 0$ mod 1 and $f'\ne 0$ mod 1.
 In this case the supercharge transforms as
 $Q_{(11)}\to \Omega_{(11)}\,Q_{(11)}$, where
  \brr \Omega_{(11)} &=&
      \exp\,(\,\pi\,f\,\Gamma_7\,\Gamma_{8}\,)\,
      \exp\,(\,\pi\,f'\,\Gamma_9\,\Gamma_{10}\,)\,\Gamma_{11} \,.
  \label{to6}\err
 We write $Q_{(11)}$ as a tensor product of a six-dimensional
 fixed-plane spinor with an $SO(5)$ ``normal" spinor according to
 $Q_{(11)}=Q_{(6)}\otimes Q_{(5)}$, and we represent the Gamma
 matrices precisely as in (\ref{gam7y}).
 In addition, we decompose the $SO(5,1)$ spinors via
 $Q_{(6)}=Q_{(6)\,R}+Q_{(6)\,L}$, where
 $=Q_{(6)\,R,L}=\pm \widehat{\Gamma}_7\,Q_{(6)\,R,L}$,
 and write the normal $SO(5)$ spinors as
 $Q_{(5)}=Q_{(4)\,R}\oplus Q_{(4)\,L}$, where
 $Q_{(4)\,R,L}$ are given by (\ref{de4}).
 Now, using the notation introduced above,
 we can write
 \brr Q_{(11)} &=&  \bpl\,Q_{(6)R} + Q_{(6)L}\,\bpr \otimes
      \bpl\,LL + LR + RL + RR \,\bpr
      \nonumber\\[.1in]
      \Omega_{(11)} &=&
      -\exp(\,i\,\pi\,f\,{\bf 1}\otimes \s_3\otimes {\bf 1}
      +i\,\pi\,f'\,{\bf 1}\otimes {\bf 1}\otimes \s_3\,)\,
      \widehat{\Gamma}_7\otimes \s_3\otimes\s_3 \,.
 \label{deco6}\err
 The fixed-plane condition $Q_{(11)}=\Omega_{(11)}\,Q_{(11)}$ can now be
 written as
 \brr Q_{(6)\,R,L}\otimes LL &=& \mp e^{-i\,\pi\,(f+f')}\,Q_{(7)}\otimes LL
      \nonumber\\[.1in]
      Q_{(6)\,R,L}\otimes LR &=& \pm e^{-i\,\pi\,(f-f')}\,Q_{(7)}\otimes LR
      \nonumber\\[.1in]
      Q_{(6)\,R,L}\otimes RL &=& \pm e^{+i\,\pi\,(f-f')}\,Q_{(7)}\otimes RL
      \nonumber\\[.1in]
      Q_{(6)\,R,L}\otimes RR &=& \mp e^{+i\,\pi\,(f+f')}\,Q_{(7)}\otimes RR
      \,.
 \err
 Therefore, as we approach a fixed-six-plane, the bulk supercharge
 is projected according to
 \brr Q_{(11)}\to \left\{\begin{array}{l}
      \,Q_{(6)L}\otimes \bpl\, LL + RR \,\bpr
      \hspace{.3in};\,\,
      f+f' \in 2\,\Z \\[.1in]
      Q_{(6)R}\otimes \bpl\, LR + RL \,\bpr
      \hspace{.3in};\,\,
      f-f'\in 2\,\Z \\[.1in]
      Q_{(6)R}\otimes \bpl\, LL + RR \,\bpr
      \hspace{.3in};\,\,
      f+f' \in 2\,\Z+1 \\[.1in]
      \,Q_{(6)L}\otimes \bpl\, LR + RL \,\bpr
      \hspace{.3in};\,\,
      f-f'\in 2\,\Z+1 \,.
      \end{array}\right. \,,
 \label{q6b}\err
 and is projected to zero if none of these conditions is met.
 Thus, we retain some supersymmetry on a fixed-six-plane
 if and only if one of the two sums $f\pm f'$ is
 integer (even or odd).  In generic cases of this sort, the bulk
 supersymmetry is {\it quartered} on the fixed-plane in question
 \footnote{A special case is when $(\,f\,,\,f'\,)=(\,1/2\,,\,1/2\,)$,
 in which case supersymmetry is merely {\it halved}.
 In that special case the difference $f-f'=0$ is an
 even integer, while the sum $f+f'=1$ is an odd integer;
 therefore, on the fixed-plane, $Q_{(11)}$ retains nonvanishing
 components of both the first and also the third case in (\ref{q6b}).}.
 Note that in the cases where the supersymmetry is
 quartered the element $\a$ necessarily has even order $N$.  In those cases
 there is necessarily an order-two element in $\Gamma$, namely $\a^{N/2}$,
 for which $(f_1,f_2,f_3\,|\,P)=(0,0,0\,|\,1)$.  The fixed-plane
 associated with $\a^{N/2}$ is ten-dimensional, while the six-plane associated
 with $\a$ is a submanifold of this ten-plane. \\

\noindent
 {\it Five-dimensional fixed-planes}\\[.1in]
 Five-dimensional fixed-planes correspond to elements
 $(f_1,f_2,f_3\,|\,P)=(f_1,f_2,f_3\,|\,0\,)$,
 where $f_{1,2,3}\ne 0$ mod 1.
 In this case, the supercharge transforms as
 $Q_{(11)}\to\Omega_{(11)}\,Q_{(11)}$, where
 \brr \Omega_{(11)}  =
      \exp(\,\pi\,f_1\,\Gamma_5\Gamma_6\,)\,
      \exp\,(\,\pi\,f_2\,\Gamma_7\Gamma_8\,)\,
      \exp\,(\,\pi\,f_3\,\Gamma_9\Gamma_{10}\,)
 \label{to5}\err
 We write $Q_{(11)}$ as a tensor product of a five-dimensional
 fixed-plane spinor with a normal spinor according to
 $Q_{(11)}=Q_{(5)}\otimes Q_{(6)}$.  Similarly, we use the
 following representation for the Gamma matrices,
 \brr \Gamma_\mu &=& \gamma_\mu\otimes {\bf 1}
      \hspace{.3in};\,\,\mu=1,...,4
      \nonumber\\[.1in]
      \Gamma_{4+i} &=& \gamma_5\otimes \widehat{\Gamma}_i
      \hspace{.3in};\,\,i=1,...,6 \,,
      \nonumber\\[.1in]
      \Gamma_{11} &=& \gamma_5\otimes \widehat{\Gamma}_7.
 \label{gam5}\err
 where $\{\,\gamma_\mu\,,\, \gamma_5\,\}$ are five-dimensional gamma
 matrices
 and $\widehat{\Gamma}_i$ are six-dimensional Gamma matrices.
 In addition, we decompose the normal $SO(6)$ spinors using the
 decomposition $SO(6)\to SO(2)\times SO(2)\times SO(2)$, where
 $SO(2)\times SO(2)\times SO(2)$ is a convenient maximal subgroup of
 $SO(6)$.  Thus, we write the normal spinor as
 $Q_{(6)}=Q_{(2)}\otimes Q_{(2)}\otimes Q_{(2)}$, and the
 six-dimensional Gamma matrices as
 $\widehat{\Gamma}_{1,2}=\s_{1,2} \otimes {\bf 1} \otimes {\bf 1}$
 and
 $\widehat{\Gamma}_{3,4}=\s_3 \otimes \s_{1,2} \otimes {\bf 1}$,
 and $\widehat{\Gamma}_{5,6}=\s_3 \otimes \s_3 \otimes \s_{1,2}$,
 where $\s_{1,2,3}$ are Pauli matrices.
 Now, using the relation
 $\widehat{\G}_7=i\,\widehat{\G}_1 \cdots\widehat{\G}_6$ we then derive
 $\widehat{\G}_7 = \s_3 \otimes \s_3 \otimes \s_3$.
 Using these expressions, it is possible to rewrite (\ref{gam5}) as
 \brr \Gamma_{1,...,4} &=& \gamma_{1,...,4}\otimes {\bf 1} \otimes {\bf 1} \otimes {\bf 1}
      \nonumber\\[.1in]
      \Gamma_{5,6} &=& \gamma_5\otimes \s_{1,2} \otimes {\bf 1} \otimes {\bf 1}
      \nonumber\\[.1in]
      \Gamma_{7,8} &=& \gamma_5\otimes \s_3 \otimes \s_{1,2} \otimes {\bf 1}
      \nonumber\\[.1in]
      \Gamma_{9,10} &=& \gamma_5\otimes \s_3 \otimes \s_3 \otimes\s_{1,2}
      \nonumber\\[.1in]
      \Gamma_{11} &=& \gamma_5\otimes \s_3 \otimes \s_3 \otimes \s_3
      \,.
 \label{gam5z}\err
 We now define
 $LLL\equiv Q_{(2)\,L}\otimes Q_{(2)\,L}\otimes Q_{(2)\,L}$, and
 similar expressions for $LLR$, $LRL$, $LRR$, $RRR$, $RRL$, $RLR$ and $RLL$,
 where $Q_{(2)\,R,L}=\pm\s_3\,Q_{(2)\,R,L}$.
 In terms of these definitions and conventions, we can write
 \brr Q_{(11)} &=& Q_{(5)}\otimes (\,RRR + RLL + LRL + LLR
      \nonumber\\[.1in]
      & & \hspace{.45in}
      + LLL + LRR + RLR + RRL\,)
      \nonumber\\[.1in]
      \Omega_{(11)} &=&
      \exp(\,i \,\pi\,f_1\,\,{\bf 1}\otimes\s_3\otimes{\bf 1}\otimes{\bf 1}
      \nonumber\\[.1in]
      & & \hspace{.2in}
      +i\,\pi\,f_2\,\,{\bf 1}\otimes\,{\bf 1}\,\otimes\s_3\otimes{\bf 1}
      \nonumber\\[.1in]
      & & \hspace{.2in}
      +i\,\pi\,f_3\,\,{\bf 1}\otimes\,{\bf 1}\,\otimes{\bf 1}\otimes\s_3\,)\,.
 \err
 The fixed-plane condition $Q_{(11)}=\Omega_{(11)}\,Q_{(11)}$ can
 now be written as
 \brr Q_{(5)}\otimes LLL &=& e^{-i\pi\,(f_1+f_2+f_3)}\,LLL
      \hspace{.5in}
      Q_{(5)}\otimes RRR = e^{+i\pi\,(f_1+f_2+f_3)}\,RRR
      \nonumber\\[.1in]
      Q_{(5)}\otimes LLR &=& e^{-i\pi\,(f_1+f_2-f_3)}\,LLR
      \hspace{.5in}
      Q_{(5)}\otimes RRL = e^{+i\pi\,(f_1+f_2-f_3)}\,RRL
      \nonumber\\[.1in]
      Q_{(5)}\otimes LRL &=& e^{-i\pi\,(f_1-f_2+f_3)}\,LRL
      \hspace{.5in}
      Q_{(5)}\otimes RLR = e^{+i\pi\,(f_1-f_2+f_3)}\,RLR
      \nonumber\\[.1in]
      Q_{(5)}\otimes LRR &=& e^{-i\pi\,(f_1-f_2-f_3)}\,RLL
      \hspace{.5in}
      Q_{(5)}\otimes RLL = e^{+i\pi\,(f_1-f_2-f_3)}\,LRR\,.
      \nonumber
 \err
 Therefore, as we approach a fixed-five-plane, the bulk
 supercharge is projected according to
 \brr Q_{(11)}\to \left\{\begin{array}{l}
      Q_{(5)}\otimes \bpl\, RRR + LLL \,\bpr
      \hspace{.4in};\,\,
      f_1+f_2+f_3 \in 2\,\Z \\[.1in]
      Q_{(5)}\otimes \bpl\, RLL + LRR \,\bpr
      \hspace{.4in};\,\,
      f_1-f_2-f_3 \in 2\,\Z \\[.1in]
      Q_{(5)}\otimes \bpl\, LRL + RLR \,\bpr
      \hspace{.4in};\,\,
      f_1-f_2+f_3 \in 2\,\Z \\[.1in]
      Q_{(5)}\otimes \bpl\, LLR + RRL \,\bpr
      \hspace{.4in};\,\,
      f_1+f_2-f_3 \in 2\,\Z
      \end{array}\right. \,,
 \err
 and is projected to zero if none of these conditions are met.
 Thus, we retain some supersymmetry on a fixed-five-plane
 if and only if at least one of the
 four sums $f_1\pm f_2\pm f_3$ is an even integer.
 In generic cases of this sort, the bulk supersymmetry is
 {\it quartered} on the fixed-plane in question. \\

\noindent
 {\it Four-dimensional fixed-planes}\\[.1in]
 Four-dimensional fixed-planes correspond to elements
 $(\,f_1\,,\,f_2\,,\,f_3\,|\,P\,)=(\,f_1\,,\,f_2\,,\,f_3\,|\,1\,)$,
 where $f_{1,2,3}\ne 0$ mod 1.
 In this case the supercharge transforms as $Q_{(11)}\to\Omega_{(11)}\,Q_{(11)}$,
 where
 \brr \Omega_{(11)} =
      \exp\,(\,\pi\,f_1\,\Gamma_5\,\Gamma_{6}\,)\,
      \exp\,(\,\pi\,f_2\,\Gamma_7\,\Gamma_{8}\,)\,
      \exp\,(\,\pi\,f_3\,\Gamma_9\,\Gamma_{10}\,)\,
      \Gamma_{11} \,.
 \err
 We write $Q_{(11)}$ as a tensor product of a four-dimensional
 fixed-plane spinor with a normal spinor according to
 $Q_{(11)}=Q_{(4)}\otimes Q_{(7)}$, and we represent the Gamma
 matrices precisely as in (\ref{gam5}).
 Furthermore, we decompose the four-dimensional spinors
 as $Q_{(4)}=Q_{(4)\,R}+Q_{(4)\,L}$ where
 $Q_{(4)\,R,L}=\pm\g_5\,Q_{(4)\,R,L}$.
 We also write the $SO(7)$ normal spinor
 as $Q_{(7)}=Q_{(6)R} \oplus Q_{(6)L}$.
 Finally, we can decompose the chiral six dimensional spinors
 $Q_{(6)R,L}$ into two dimensional chiral spinors as descried above.
 Using these conventions, we can write
 \brr Q_{(11)} &=& (\,Q_{(4)\,R}+Q_{(4)\,L}\,) \otimes
      (\,RRR + RLL + LRL + LLR
      \nonumber\\[.1in]
      & & \hspace{1.4in}
      + LLL + LRR + RLR + RRL \,)
      \nonumber\\[.1in]
      \Omega_{(11)} &=&
      \exp\bpl\,i\,\pi\,f_1\,\,{\bf 1}\otimes \s_3 \otimes\,{\bf 1}\,\otimes {\bf 1}
      \nonumber\\[.1in]
      & & \hspace{.3in}
      +i\,\pi\,f_2\,\,{\bf 1}\otimes\,{\bf 1}\,\otimes \s_3 \otimes {\bf 1}
      \nonumber\\[.1in]
      & & \hspace{.3in}
      +i\,\pi\,f_3\,\,{\bf 1}\otimes\,{\bf 1}\,\otimes\,{\bf 1}\,\otimes \s_3
      \,\bpr\,
      \g_5\otimes\s_3\otimes\s_3\otimes\s_3\,.
 \label{q11q4q4} \err
 The fixed-plane condition $Q_{(11)}=\Omega_{(11)}\,Q_{(11)}$ can
 now be written as
  \brr Q_{(4)\,R,L}\otimes LLL &=& \mp e^{-i\pi\,(f_1+f_2+f_3)}\,LLL
      \hspace{.3in}
      Q_{(4)\,R,L}\otimes RRR = \pm e^{+i\pi\,(f_1+f_2+f_3)}\,RRR
      \nonumber\\[.1in]
      Q_{(4)\,R,L}\otimes LLR &=& \pm e^{-i\pi\,(f_1+f_2-f_3)}\,LLR
      \hspace{.3in}
      Q_{(4)\,R,L}\otimes RRL = \mp e^{+i\pi\,(f_1+f_2-f_3)}\,RRL
      \nonumber\\[.1in]
      Q_{(4)\,R,L}\otimes LRL &=& \pm e^{-i\pi\,(f_1-f_2+f_3)}\,LRL
      \hspace{.3in}
      Q_{(4)\,R,L}\otimes RLR = \mp e^{+i\pi\,(f_1-f_2+f_3)}\,RLR
      \nonumber\\[.1in]
      Q_{(4)\,R,L}\otimes LRR &=& \mp e^{-i\pi\,(f_1-f_2-f_3)}\,RLL
      \hspace{.3in}
      Q_{(4)\,R,L}\otimes RLL = \pm e^{+i\pi\,(f_1-f_2-f_3)}\,LRR
      \nonumber
 \err
 Therefore, as we approach a fixed-four-plane, the bulk
 supercharge is projected according to
 \brr Q_{(11)}\to \left\{\begin{array}{l}
      \,Q_{(4)L}\otimes LLL \, + \,Q_{(4)R}\otimes RRR
      \hspace{.4in};\,\,
      f_1+f_2+f_3\in 2\,\Z \\[.1in]
      \,Q_{(4)R}\otimes LRL \, + \,Q_{(4)L}\otimes RLR
      \hspace{.4in};\,\,
      f_1-f_2+f_3\in 2\,\Z \\[.1in]
      \,Q_{(4)R}\otimes RRL \, + \,Q_{(4)L}\otimes LLR
      \hspace{.4in};\,\,
      f_1+f_2-f_3\in 2\,\Z \\[.1in]
      \,Q_{(4)L}\otimes RLL \, + \,Q_{(4)R}\otimes LRR
      \hspace{.4in};\,\,
      f_1-f_2-f_3\in 2\,\Z \\[.1in]
      \,Q_{(4)R}\otimes LLL \, + \,Q_{(4)L}\otimes RRR
      \hspace{.4in};\,\,
      f_1+f_2+f_3\in 2\,\Z+1\\[.1in]
      \,Q_{(4)L}\otimes LRL \, + \,Q_{(4)R}\otimes RLR
      \hspace{.4in};\,\,
      f_1-f_2+f_3\in 2\,\Z+1 \\[.1in]
      \,Q_{(4)L}\otimes RRL \, + \,Q_{(4)R}\otimes LLR
      \hspace{.4in};\,\,
      f_1+f_2-f_3\in 2\,\Z+1 \\[.1in]
      \,Q_{(4)R}\otimes RLL \, + \,Q_{(4)L}\otimes LRR
      \hspace{.4in};\,\,
      f_1-f_2-f_3\in 2\,\Z+1 \\[.1in]
      \end{array}\right. \,,
 \label{res4d}\err
 and is projected to zero if none of these conditions are met.
 Thus, we retain some supersymmetry on a fixed-four-plane
 if and only if any one of the four sums
 $f_1\pm f_2\pm f_3$ is an integer (even or odd).
 In generic cases of this sort, supersymmetry is {\it eighthed}
 on the fixed-plane in question.

 There are special cases, however.  For instance, for each of
 the two cases where (\,$f_1$\,,\,$f_2$\,,\,$f_3$\,) are given by
 (\,$1/2$\,,\,$1/3$\,,\,$1/6$\,) and by
 (\,$1/2$\,,\,$1/4$\,,\,$1/4$\,), one sum
 $f_1-f_2-f_3=0$ is an even integer, while
 another sum $f_1+f_2+f_3=1$ is an odd integer.
 Therefore, on the fixed-plane, $Q_{(11)}$ retains
 nonvanishing components of both the fourth and the
 also the fifth case in (\ref{res4d}).  Thus, supersymmetry
 is merely quartered to $D=4, N=2$, rather than eighthed
 to $D=4, N=1$, in these cases
 \footnote{If the fixed-four-plane in question is a submanifold
 of the fixed-plane of another element of $\Gamma$,
 one needs to account for the action of this
 other element as well before drawing conclusions
 as to the amount of fixed-plane supersymmetry.}.

 Note that the pairs of summands indicated in
 in (\ref{res4d}) have correlations owing to the Majorana
 constraint on $Q_{(11)}$.  As a result, the two apparent
 summands in each of these expressions actually describe
 the same degrees of freedom.
 In terms of the choices we have made, this gives rise to a Majorana
 spinor supercharge in four dimensions.  The pairs which
 appear in (\ref{res4d}) represent left- and right-handed
 chiral projections of this four-dimensional Majorana supercharge.
 As is well known, in four dimensions spinor fields which are
 Majorana may be equally well represented in terms of either
 chiral projection.

 \setcounter{equation}{0}
 \section{Generalization to Non-Abelian Orbifold Groups}
 \label{nonabelian}

 One obvious extension of our criterion, still for orbifolds of
tori of our class, is to consider {\it all} the automorphisms of
the underlying lattices. In addition to the rotations and flips we
have already considered, the only new type of ``generating''
automorphism to deal with is the permutations of the coordinates
$x^i$.  In fact, one can easily lift these permutation actions to
spinors. Here are the details.

 The lattices of our class are all of the form
 \begin{equation}
 \Lambda=(A_1)^a\oplus(A_2)^b \ , \label{lattype}
 \end{equation}
 with $a + 2 b = n$, the dimension of the orbifold. Such a lattice
 has an automorphism group of order
 $2^a \cdot a! \cdot b! \cdot 12^b$\,,
 a product of four factors.  Let's check where these two pairs of
 factors come from.

 The first pair, $2^a \cdot a!$\,, keeps track of the sign flips on
 each of $a$ possible $x^i$ (the $2^a$), and permutations of the
 same $a$ $x^i$'s (the $a!$).  We have already described a way to
 lift the flips to spinors, via $\G_i$, so we only need to lift the
 permutations. The second pair, $b! \cdot 12^b$, keeps track of the
 permutations of the $b$ different copies of $A_2$, i.e., of
 corresponding ordered pairs $(x^i, x^{i+1})$ (giving the $b!$),
 and automorphisms internal to each of the $b$ $A_2$'s (giving the
 $12^b$, as the automorphism group of the $A_2$ lattice is the
 dihedral group $D_6$ of order $12$).  Consider a single $A_2$
 lattice in the ``$(x,y)$-plane''.  The dihedral automorphism group
 $D_6$ consists entirely of the identity, rotations by $r^k :=  2
 \pi k / 6$, $k = 1, 2, 3, 4, 5$, and flips ($s, s r, s r^2 , s r^3
 , s r^4 , s r^5$) across lines through opposite vertices or
 opposite edge centers of the period hexagon (we can assume that
 $s$ is the flip across the $x$-axis sending $y \mapsto -y$).  In
 particular, $D_6$ is generated by a rotation $r$ and a flip $s$,
 for each of which we have already described a lift to spinors.
 What remains is again to describe the effect of the permutation
 automorphisms on spinors.

 Since every permutation of the $x^i$ can be expressed as a product
 of simple transpositions $x^i \leftrightarrow x^j$, $i \neq j$, it
 suffices to write out the lift of such a transposition.  Imagine
 the $(x^i,x^j)$-plane.  Rotation counterclockwise by $\pi /2$
 radians sends $x^i \mapsto x^j$ and $x^j \mapsto - x^i$.  This is
 represented in the Clifford algebra by
 \brr \exp (\,\frac{\pi}{4}\,\G_{ij}\,) = \cos(\,\pi / 4\,) + \sin(\,\pi / 4\,)\,\G_i \G_j =
      \left( \frac{\sqrt{2}}{2} + \frac{\sqrt{2}}{2}\,\G_i \G_j \right) \,.
 \err
 Now compose this with the flip sending $x^i \mapsto - x^i$ given
 by $\G_i$, yielding
 \brr \G_i \cdot \left( \frac{\sqrt{2}}{2} + \frac{\sqrt{2}}{2} \G_i \G_j \right)
      =  \frac{\sqrt{2}}{2}\,(\,\G_i + \G_j\,) \,.
 \err
 We can build any permutation we like out of a product of these,
 and in particular all of the remaining automorphisms come from
 mixing these with the rotations and flips as above.
 By adding such permutations, we can obtain a complete
 supersymmetric restriction criterion for all orbifolds modelled on
 lattices of type (\ref{lattype}).   In particular, this allows one
 to check the supersymmetry constraints for a large class of
 {\it{nonabelian}} orbifold groups. The generalized supersymmetric
 restriction theorem, including the extension to non-abelian
 orbifold groups, will be discussed in a forthcoming paper.

 \setcounter{equation}{0}
 \section{Soft Orbifolds and $G_2$-structures}

 \label{softsect}
 Consider the supersymmetric orbifold of $\Gamma =
 ({\mathbb{Z}}_2)^3$ labelled $(1111111)$ in Table \ref{scan2}. The
 corresponding representation of $\Gamma$ on ${\mathbb{R}}^7$ is
 given by the direct sum of the seven nontrivial characters
 $\Gamma_{[i,j,k]}$, each with multiplicity one.  The character
 table for $\Gamma = \langle \alpha, \beta, \gamma \rangle$ is
 given in Table \ref{charz2cube}, and, by following the prescription given
 in Section \ref{orbenum}, we see that the action of these generators on the
coordinate $7$-tuple $(x_1, \ldots, x_7)$ is given by \brr
\alpha & : & (x_1, x_2, x_3, -x_4, -x_5, -x_6, -x_7) \\
\beta & : & (x_1, -x_2, -x_3, x_4, x_5, -x_6, -x_7) \\
\gamma & : & (-x_1, x_2, -x_3, x_4, -x_5, x_6, -x_7) \ . \err
 \begin{table}
 \begin{center}
 \begin{tabular}{|c|cc|cc|cc|cc|}
 \hline
 &&&&&&&&\\[-.1in]
 $(\Z_2)^3$ & $\Omega$ & $\Gamma_{[0,0,1]}$ & $\Gamma_{[0,1,0]}$
 & $\Gamma_{[0,1,1]}$ & $\Gamma_{[1,0,0]}$ & $\Gamma_{[1,0,1]}$ & $\Gamma_{[1,1,0]}$ & $\Gamma_{[1,1,1]}$ \\[.1in]
 \hline
 &&&&&&&& \\[-.1in]
 $1$ & + & + & + & + & + & + & + & + \\[.1in]
 $\gamma$ & + & $-$ & + & $-$ & + & $-$ & + & $-$ \\[.1in]
 \cline{2-9}
 &&&&&&&&\\[-.1in]
 $\beta$ & + & + & $-$ & $-$ & + & + & $-$ & $-$ \\[.1in]
$\beta \gamma$ & + & $-$ & $-$ & + & + & $-$ & $-$ & + \\[.1in]
 \cline{2-9}
 &&&&&&&&\\[-.1in]
 $\alpha$ & + & + & + & + & $-$ & $-$ & $-$ & $-$ \\[.1in]
 $\alpha \gamma$ & + & $-$ & + & $-$ & $-$ & + & $-$ & +  \\[.1in]
 \cline{2-9}
 &&&&&&&&\\[-.1in]
 $\alpha \beta$ & + & + & $-$ & $-$ & $-$ & $-$ & + & + \\[.1in]
 $\alpha \beta \gamma$ & + & $-$ & $-$ & + & $-$ & + & + & $-$ \\[.1in]
 \hline
 \end{tabular} \\[.2in]
 \caption{Character table for the group $(\Z_2)^3$.}
 \label{charz2cube}
 \end{center}
 \end{table}

Note that this $\Gamma$-action preserves the $G_2$-invariant
differential $3$-form $\varphi_0$ on ${\mathbb{R}}^7$,
$$\varphi_0 := dx_{123} + dx_{145}
+ dx_{167} + dx_{246} - dx_{257} - dx_{347} - dx_{356} \ ,$$ where
$dx_{ijk} := dx_i \wedge dx_j \wedge dx_k$.  Such an action
defines a {\it $G_2$-structure} on the orbifold $T^7 / \Gamma$.

Our supersymmetric hard orbifold $(1111111)$ is one of a family of
orbifolds considered by Joyce\footnote{In fact, in the notation of
\cite[Equations (23)-(25)]{JoyJDG2}, ours corresponds to $b_1 =
b_2 = c_1 = c_3 = c_5 = 0$.  We will refer to notation and
examples from his book \cite{Joyce} instead.}. It was his goal to
construct compact manifolds with holonomy group $G_2$. To this end
he developed in \cite{JoyJDG1, JoyJDG2, Joyce} a machinery which,
starting with a sufficiently simple orbifold admitting a
$G_2$-structure, establishes the existence of such a metric on a
``resolution" of the orbifold.  Joyce's method depends on the
existence of certain {\it R-data} (``R" for Resolution) which, for
a given orbifold with $G_2$-structure, may yield a large number of
topologically distinct $G_2$ holonomy manifolds as resolutions.

In \cite[Sections 12.2,\,3,\,\&\,5]{Joyce} three paricular
softenings are considered.  The actions of the generators of
$\Gamma$ take the form \brr
\alpha & : & (x_1, x_2, x_3, -x_4, -x_5, -x_6, -x_7) \\
\beta & : & (x_1, -x_2, -x_3, x_4, x_5, b - x_6, -x_7) \\
\gamma & : & (-x_1, x_2, -x_3, x_4, c - x_5, x_6, \frac{1}{2} -
x_7) \ , \err with $b, c \in \left\{0, \frac{1}{2}\right\}$. In
fact, Joyce considers the three cases
$$ (b, c) = \left(0, \frac{1}{2}\right) \, , \ \left(\frac{1}{2}, 0\right) \, , \
\mbox{and} \ \left(\frac{1}{2}, \frac{1}{2}\right) \, ,$$ and
shows that each of these admits a set of R-data  defining a $G_2$
holonomy resolution.  In each case the generators $\alpha$,
$\beta$, and $\gamma$ act with fixed-points as before, but some
other elements, like \brr  \beta \gamma & : & (-x_1, -x_2, x_3,
x_4, c - x_5, b + x_6, \frac{1}{2} + x_7) \ , \err act freely.
Thus softening an orbifold has the effect of simplifying the
singularities.  The singularities of the orbifold $(1111111)$ are
too complicated to visibly admit a compatible set of R-data, hence
his interest in various softenings with simpler singularities.
Furthermore, at least for these simplest examples, the set of
fixed-plane dimensions associated to a hard orbifold contains as
subsets those of all of its softenings. For this reason, one
expects that much of the supersymmetric restriction analysis of
Section \ref{theorem} can be usefully adapted to the case of soft
orbifolds.

It is clear that having a $G_2$-structure on a hard orbifold
implies such a structure on all of its softenings.  This raises
the intriguing question of the relationship between our
supersymmetry condition and such $G_2$-structures: Do all
supersymmetric hard orbifolds (and hence all of their softenings)
carry a $G_2$-structure? What about the converse: Do all orbifolds
with a $G_2$-structure arise via softening from supersymmetric
hard orbifolds?  What about the more subtle issue of resolvability
as a $G_2$-manifold (i.e., existence of compatible R-data)?  These
are all questions to be addressed in future work.

 \setcounter{equation}{0}
 \section{Conclusions}
 \label{extsect}

We have described a systematic method for classifying
supersymmetric orbifold compactifications of {\it M}-theory,
specifically hard orbifolds defined by pseudo-planar
representations of abelian groups of order $\leq 12$. Although we
stopped our ``periodic table" (Table \ref{scan2}) at order $12$,
the methods developed here apply to groups of arbitrary order, and
the algorithmics are such that such an extension (using
Mathematica) is computationally feasible.

 It is physically relevant that we demand
 orbifold actions be compatible with particular lattices.
 By doing this we are able to keep control over two separate
 problems.  On the one hand we are studying aspects of
 supersymmetric singularities. At the same time, however,
 we are learning something about which sets
 of such singularities, and their neighborhoods, can be assembled
 together to create a global supersymmetric compactification
 space.  Elucidating the relationship between our own supersymmetric
 configurations of singularities and those of intersecting branes as studied by
 other groups
 \cite{bkl1, bkl2, BBKL, AchC, AchD, AchE, afiru1, afiru2, imr, csu, csu1, csu2}
 seems an important direction for further investigation.

 Furthermore, we touch upon the interesting
 mathematical problem of classifying $G_2$ manifolds.  The vast
 majority of known compact $G_2$ holonomy manifolds arise by Joyce's resolution
 of singularities from  orbifolds of $T^7$.  Physically
 one expects
 that seven dimensional supersymmetric compactification spaces of
 the sort we are studying should admit such $G_2$ resolutions.  In
 order to compare in general Joyce's ``resolution data" to the
 constraints arising from our global supersymmetric restriction, it
 seems necessary to formulate the restriction for the soft
 orbifolds of the introduction, as all of Joyce's examples
 are of this sort.

 \appendix

 \section{The $C$-Matrices for Selected Groups}
 \label{cmatlist}
 By applying the simple algorithm described in section
 \ref{cmatrices}, we are able to derive the $C$-matrix for any
 pseudo-planar group $\Gamma=G_1\times...\times G_n$,
 where $G_i\in\{\,\Z_2\,,\,\Z_3\,,\,\Z_4\,\}$ for $1 \leq i \leq n$.  In
 Tables 7 and 8 we exhibit these for all cases for
 which $|\,\Gamma\,|\le 12$.
 \begin{center}
 \brr && \nonumber\\[1in]
      C(\Z_2) &=& 1
      \nonumber\\[.1in]
      C(\Z_3) &=& 1
      \nonumber\\[.1in]
      C(\Z_4) &=& \ba{c|c}0&2\\\hline 2&1\ea
      \nonumber\\[.1in]
      C(\Z_2\times \Z_2) &=&
      \ba{ccc}
      0&2&2\\
      2&0&2\\
      2&2&0\ea
      \nonumber\\[.1in]
      C(\Z_2\times\Z_3) &=& \ba{c|c|c}3&0&3\\\hline 0&2&2\\\hline 3&2&-1\ea
      \nonumber\\[.1in]
      C(\Z_2\times\Z_4) &=&
      \ba{ccc|cc}0&0&0&4&4\\0&4&4&0&4\\0&4&4&4&0\\\hline 4&0&4&2&2\\4&4&0&2&-2\ea
      \nonumber\\[.1in]
      C(\Z_2\times\Z_2\times\Z_2) &=&
      \ba{ccccccc}
      4&0&4&0&4&0&4\\
      0&4&4&0&0&4&4\\
      4&4&0&0&4&4&0\\
      0&0&0&4&4&4&4\\
      4&0&4&4&0&4&0\\
      0&4&4&4&4&0&0\\
      4&4&0&4&0&0&4\ea
      \nonumber
 \err \\[.2in]
 Table 7: The $C$-matrices for pseudoplanar abelian groups of order $\leq 8$.
 \label{cless8}
 \end{center}

 \begin{center}
 \brr C(\Z_3\times\Z_3) &=&
      \ba{cccc}0&3&3&-3\\3&0&3&3\\3&3&-3&0\\-3&3&0&3\ea
      \nonumber\\[.1in]
      C(\Z_2\times\Z_2\times\Z_3) &=&
      \ba{ccc|c|ccc}
      0&6&6&0&0&6&6\\
      6&0&6&0&6&0&6\\
      6&6&0&0&6&6&0\\
      \hline
      0&0&0&4&4&4&4\\
      \hline
      0&6&6&4&4&-2&-2\\
      6&0&6&4&-2&4&-2\\
      6&6&0&4&-2&-2&4\ea
      \nonumber\\[.1in]
      C(\Z_3\times\Z_4) &=&
      \ba{c|c|c|c|cc}
      0&0&6&0&6&6\\\hline
      0&4&0&4&4&4\\\hline
      6&0&3&6&3&-3\\\hline
      0&4&6&4&-2&-2\\\hline
      6&4&3&-2&-5&1\\
      6&4&-3&-2&1&-5\ea
      \nonumber
 \err \\[.2in]
 Table 8: The $C$-matrices for pseudoplanar abelian groups of orders $9$ and $12$.
 \label{cbet912}
 \end{center}

 \vspace{.5in}
 {\Large {\bf Acknowledgements}}\\[.1in]
 C.F.D. would like to thank P.~Candelas, N.~Hitchin, and X.~de la Ossa of the
 Mathematics Institute, Oxford for interesting and encouraging
 discussions. M.F. is grateful to M{\'a}ria and Emil Martinka,
 for hospitality, encouragement and halu{\v s}ky at the Slovak
 Institute for Basic Research, Podvazie Slovakia, where some of
 this manuscript was prepared.

 \end{document}